\documentclass[%
reprint,
superscriptaddress,
 aps,
 pra,
longbibliography,
dvipsnames,
twocolumn,
xcolor=table
]{revtex4-1}
\usepackage[main=english]{babel}
\usepackage{amssymb}

\usepackage{dcolumn}% Align table columns on decimal point
\usepackage{siunitx}
\usepackage{dsfont}

\usepackage[utf8]{inputenc}
\usepackage[toc,page]{appendix}
\usepackage{graphicx}
\usepackage{hyperref}
\hypersetup{colorlinks}
\usepackage{soul}

\usepackage{physics}
\usepackage{bbold}
\usepackage{float}
\usepackage{hyphenat}

\usepackage{qcircuit}
\usepackage{amssymb}

\begin{document}
\newcommand{\orcidicon}[1]{\href{https://orcid.org/#1}{\includegraphics[height=\fontcharht\font`\B]{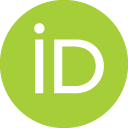}}}
\hypersetup{colorlinks=true,citecolor=magenta,linkcolor=magenta,filecolor=magenta,urlcolor=magenta,breaklinks=true}

\newcommand{\Obs}{\hat{O}(\vec{q})}
\newcommand{\TE}[1]{\mathcal{U}(#1)}
\author{Oriel~Kiss\orcidicon{0000-0001-7461-3342}}
\email{oriel.kiss@cern.ch}
\affiliation{European Organization for Nuclear Research (CERN), Geneva 1211, Switzerland}
\affiliation{Department of Nuclear and Particle Physics, University of Geneva, Geneva 1211, Switzerland}
\affiliation{Physics Department, University of Trento, Via Sommarive 14, I-38123 Trento, Italy}

\author{Michele~Grossi\orcidicon{0000-0003-1718-1314}}
\affiliation{European Organization for Nuclear Research (CERN), Geneva 1211, Switzerland}

\author{Alessandro~Roggero\orcidicon{0000-0002-8334-1120}}
\affiliation{Physics Department, University of Trento, Via Sommarive 14, I-38123 Trento, Italy}
\affiliation{INFN-TIFPA Trento Institute of Fundamental Physics and Applications, Trento, Italy}

\title{Quantum error mitigation for Fourier moment computation}

\date{\today}

\begin{abstract}
Hamiltonian moments in Fourier space - expectation values of the unitary evolution operator under a Hamiltonian at different times - provide a convenient framework to understand quantum systems. They offer insights into the energy distribution, higher-order dynamics, response functions, correlation information and physical properties. This paper focuses on the computation of Fourier moments within the context of a nuclear effective field theory on superconducting quantum hardware. The study integrates echo verification and noise renormalization into Hadamard tests using control reversal gates. These techniques, combined with purification and error suppression methods, effectively address quantum hardware decoherence. The analysis, conducted using noise models, reveals a significant reduction in noise strength by two orders of magnitude. Moreover, quantum circuits involving up to 266 CNOT gates over five qubits demonstrate high accuracy under these methodologies when run on IBM superconducting quantum devices. 
\end{abstract}

\maketitle

\section{Introduction}
 Quantum computers offer a natural paradigm for Hamiltonian simulations, with numerous applications in nuclear~\cite{Dumitrescu_2018,roggero2019} and condensed matter physics \cite{Hofstetter_2018,quantum_simulations_XXZ}, quantum field theory \cite{QS_QFT_Preskill, QS_Schwinger_Lougovski,Klco_2022,QS_lattice_Muschnik,farrell2023scalable} and quantum chemistry \cite{Su2021nearlytight, Ouyang2020compilation, Martinez_partitioning}. The backbone of numerous algorithms is given by the ability to simulate real\hyp time evolution efficiently. For instance, these algorithms have applications such as the computation of energy levels in chemistry via quantum phase estimation \cite{QPE-Lloyd, Arrazola_batteries,PRXQ_arrazola}, prediction of chemical reaction rates \cite{reaction}, correlation functions \cite{two_point_roggero,Hofstetter_2018,quantum_simulations_XXZ,nature_grossi}, neutrino oscillations \cite{PRD_neutrino,illa2022anneal,neutrino_simulation_amitrano,illa2022multi} and scattering experiments \cite{neutrio_nucleus_roggero, du2021}. Quantum dynamics with more than a few particles become quickly overwhelming for classical devices, making these problems promising early applications of quantum computers.

 Response functions, which describe the linear response of a many\hyp body system after an excitation, are generally challenging to compute from first principles, making them an appealing application for quantum computers, see e.g. \cite{review_standard_model_roggero} for a review. For instance, they can describe scattering processes by probing the internal structure of the target \cite{dyn_lin_response}, as they contain the same information as an inclusive reaction cross\hyp section. They are typically expressed in terms of the spectral density operator $\delta(\omega-(H-E_0))$, where $E_0$ is the target ground state energy and $H$ the Hamiltonian describing the target. The coupling of the target to the external probe is described by an excitation operator $\hat{O}(\vec{q})$, depending on the momentum transfer $\vec{q}$ of the scattering process. At high momenta, the response functions are expected to be mainly dependent on the target's momentum distribution or spectral functions. However, at more modest momentum transfer, two\hyp body interactions are essential when measuring neutrino properties through neutrino\hyp nuclei experiments, including  MiniBooNE, MicroBooNE, T2K, and DUNE \cite{neutrino_experiments}.
 
 Given an Hamiltonian $H$, an initial state $|\Psi_0\rangle$ for the target and an excitation operator $\hat{O}(\vec{q})$, the frequency dependant \textit{response function} is defined as
\begin{equation}
\label{eq:S}
    \begin{split}
        S(\omega,\vec{q}) &= \langle \Psi_0| \Obs^\dagger \delta(\omega-(H-E_0))\Obs|\Psi_0\rangle \\
        &=\sum_f \left| \langle \Psi_0|\Obs|f\rangle \right|^2\delta(\omega-(E_f-E_0)),
    \end{split}
\end{equation}
where $\{|f\rangle\}_f$ are the eigenstates of $H$ with corresponding energies $\{E_f\}_f$. Computing this value is generally extremely challenging in practice since it requires knowledge of the full spectrum. A useful strategy to circumvent this problem is to consider instead an integral transform
\begin{equation}
\begin{split}
\label{eq:int_transf}
\Phi(\nu,\vec{q})&=\int d\omega K(\nu,\omega)S(\omega,\vec{q})\\
&=\langle \Psi_0| \Obs^\dagger K(\nu,(H-E_0))\Obs|\Psi_0\rangle
\end{split}
\end{equation}
which, for suitable kernels, allows for a direct calculation using ground state techniques. For instance, in Quantum Monte Carlo calculations the Laplace kernel is typically used thanks to the relation between $\Phi(\nu)$ in that case and imaginary-time correlation functions~\cite{Carlson1992,Carlson2015,Lovato2016,neutrino_experiments}. Another common choice, preferred in many-body methods like Coupled Cluster or exact diagonalization, is the Lorentz Integral Transform~\cite{EFROS1994130,Efros_2007,Bacca2013,Bacca2014,Sobczyk2021}.

The integral transform $\Phi(\nu,\vec{q})$ can be thought as a generalization of energy moments~\cite{moment} providing valuable information about the response function while being more accessible. For sufficiently smooth kernels one can expand $K(\nu,\omega)$ onto a basis of orthogonal polynomials reducing the calculation of the response function into the calculation of the expansion coefficients in the chosen basis. In the present work we focus on the computation of Fourier moments which are the coefficients of the expansion of the response function into a basis of plane waves. Thanks to the expected ability of quantum simulations to provide access to the real-time evolution of many-body systems, the Fourier basis has already been suggested as a tool to obtain spectral properties with efficient quantum algorithms~\cite{L&T,somma2019quantum,Lu2021,randomized_QPE} and, for nuclear physics application, it has been shown to also allow for cheaper calculations by incorporating directly additional available information about the response function~\cite{faster_spectral_roggero}.

This paper focuses on the computation of the Fourier moments for the response function describing inelastic scattering between a Triton and a lepton on noisy intermediate scale quantum (NISQ) devices \cite{Preskill2018quantumcomputingin}. We adapt two error mitigation techniques, echo verification (EV) \cite{OBrien_verification} and operator decoherence renormalization (ODR) \cite{mitigating_depolarising,self-mitigation, farrell2023scalable} to Hadamard tests with and without control reversal gates. The performance of IBM NISQ devices is enhanced using different error suppression techniques, such as dynamical decoupling \cite{DD_Lloyd}, pulse\hyp efficient transpilation \cite{pulse-efficient} and randomized compiling \cite{Randomized_compiling, Pauli_Twirling}.

The computation of response functions from Fourier moments is explained in Sec.~\ref{response_sec}. Sec.~\ref{hadamard_test} describes the error mitigation techniques for standard Hadamard tests while Sec.~\ref{CRG} handles the control\hyp free setting using control reversal gates. The physical system is introduced in Sec.~\ref{physical_model}, and the results are displayed in Sec.~\ref{res_sec}. Additional details about error suppression, confidence intervals calculation techniques and explicit circuit construction are given in the appendices. 

\section{Methods}
In this section, we present the main tools of this work: how to compute the response function from Fourier moments and the implementation of echo verification and operator decoherence renormalization with and without control reversal gates. 
\subsection{Response function on a quantum computer}
\label{response_sec}
As we discussed in the introduction, a direct computation of the response function, see Eq.~\eqref{eq:S}, is prohibitive in general as it requires access to the full spectrum while a suitable integral transform, see Eq.~\eqref{eq:int_transf}, could be estimated by measuring directly properties of the groundstate. For the integral transform to be useful in reconstructing the energy dependence of the response function with minimal uncontrollable errors one typically requires integral kernels that are smooth approximations of the energy delta function from Eq.~\eqref{eq:S} with a finite width (see~\cite{PRA_roggero_gaussian_integral} for a more detailed characterization) such as a Gaussian or Lorentzian function. By taking a translationally invariant integral kernel, i.e. where $K(\nu,\omega)=K(|\nu-\omega|)$, and  expressing it in terms of a set of orthogonal polynomials $\phi_j$ we have
\begin{equation}
\Phi(\nu) = \sum_j^{\infty} c_j(\nu) \langle\Psi_0\lvert\Obs^\dagger \phi_j(H)\Obs\rvert\Psi_0 \rangle\;,
\end{equation}
where the coefficients $c_j(\nu)$ specify the particular kernel function employed.
In order for this procedure to be efficient, it should be possible to truncate the series at a finite number of terms with a suitably small error. For instance, a Gaussian kernel expressed in terms of either Chebyshev or Fourier orthogonal polynomials satisfy this requirement~\cite{PRA_roggero_gaussian_integral,PRE_roggero_chebychev,faster_spectral_roggero}. Using the Fourier basis and truncating the series expansion up to orders $|j|\leq N$, the integral transform can be written as

\begin{equation}
\label{eq:response}
\Phi_N (\nu) = \sum_{j=-N}^{N} c_j(\nu) m(j\tau),
\end{equation}
where 
\begin{equation}
\label{eq_moment}
\begin{split}
m(j\tau)=&\left<\Psi_0\right|\Obs^\dagger e^{-iHj\tau}\Obs\left|\Psi_0 \right>
\end{split}
\end{equation}
are the Fourier moments meant to be computed on a quantum computer. The choice of the time-step $\tau$ and the truncation order $N$ affect the minimum energy resolution $\Delta$ that can be achieved with the integral transform for a fixed accuracy. In particular one can show that, with the Gaussian kernel, the total evolution time $T=N\tau$ needs to scale as $T=\mathcal{O}(1/\Delta\sqrt{\log(1/\epsilon)})$ in order to keep the approximation error below $\epsilon$ (for more details see~\cite{PRA_roggero_gaussian_integral,faster_spectral_roggero}).

\subsection{Real time evolution}
\label{time_evolutionm_sec}
Given a Hamiltonian $H=\sum_\gamma^\Gamma H_\gamma$ composed of $\Gamma$ summands, we are interested in the real\hyp time evolution operator $\mathcal{U}(t)$ generated by $H$, given by $\exp{-it\sum_\gamma ^\Gamma H_\gamma}$. Product formulas, such as first\hyp order Trotter 
\begin{equation} 
\mathcal{U}_1(t) = \prod_\gamma^\rightarrow e^{-itH_\gamma} = \mathcal{U}(t) +\mathcal{O}(t^2),
\end{equation} 
second\hyp order Trotter\hyp Suzuki \cite{Suzuki,SUZUKI1990319}
\begin{equation} 
\mathcal{U}_2(t) = \prod_\gamma^\rightarrow e^{-it/2H_\gamma}\prod_\gamma^\leftarrow e^{-it/2H_\gamma} = \mathcal{U}(t) +\mathcal{O}(t^3),
\end{equation} 
and higher\hyp order products
 \begin{equation}
 \begin{split}
     \mathcal{U}_{2j}(t) &= \mathcal{U}_{2j-2}(p_jt)^2\mathcal{U}_{2j-2}((1-4p_j)t) \mathcal{U}_{2j-2}(p_jt)^2\\ 
     &= \mathcal{U}(t)+\mathcal{O}(t^{2j+1})\\
     p_j &= (4-4^{1/(2j-1)})^{-1},
     \end{split}
 \end{equation}
 are popular methods which break the matrix exponential into a product of simpler terms with controllable error. We note that the higher\hyp order product formulas are recursively defined and that the arrow over the product sign determines the order in which the factors have to be multiplied together. The error scaling $\mathcal{O}(t^{j+1})$, with $j$ being the order of the product, can be further improved by splitting the evolution time $t$ into $r$ smaller time\hyp slice, better known as Trotter steps, leading to 
 \begin{equation}
    \mathcal{U}_{2j}\left(\frac{t}{r}\right)^r = \mathcal{U}(t) + \mathcal{O}\left(\frac{t^{2j+1}}{r^{2j}}\right).
 \end{equation}
Even if better error estimates have been obtained~\cite{PhysRevX_high_trotter} by taking into account the commutators $\sum_{\gamma,\gamma'} [H_\gamma,H_{\gamma'}]$, product formulas are usually more accurate than these theoretical bounds. In fact, computing tighter error bound remains an active area of research. For our purposes, it remains preferable to empirically estimate the performance, e.g. by increasing the number of steps until the improvement becomes smaller than some tolerance. We guide the reader to \cite{simulation_tacchino, quantum_simulations_lanes} for informative reviews about actual implementation.

Many proposal have been made to improve the accuracy of product formulae. Randomization, e.g. in the product ordering \cite{Childs2019fasterquantum,doubling_random_berry} or in the time splitting \cite{Faehrmann2022randomizingmulti} has proven to be the most common denominator of these approaches. These tools effectively increase the order of the product formulae, and thus boost their performance. Moreover, random compilers, such as qDRIFT \cite{QDrift} and studied further in \cite{Ouyang2020compilation,QDRift_caltech,qDRIFT_Kiss,nakaji2023qswift}, take a step further by sampling product formulae directly from the Hamiltonian. Such random products are therefore strong candidates when a large number of terms $\Gamma$ are present, and when the distribution of the coefficients is non-uniform. Alternatively, more refined techniques do exist, using extra ancillary qubits and complex gadgets, such as qubitization and linear combination of unitaries \cite{Child,Black_box_Berry, Low2019hamiltonian,PhysRevX.8.041015}. They usually offer better asymptotic scaling but are more challenging to implement in practice, as they generally require fault\hyp tolerant devices, and are outside the scope of this paper.

\subsection{Standard Hadamard test}
\label{hadamard_test}
The direct evaluation of the Fourier moments $m(j\tau)$ from Eq.~\eqref{eq_moment} on a quantum computer becomes more challenging in the common situation where the excitation operators $\Obs$ is not unitary. Without loss of generality we can always expand them as
\begin{equation}
\Obs=\sum_{k=1}^{N_O}o_kO_k\;,
\end{equation}
with $o_k\in\mathbb{R}$ and $O_k$ unitary operators. Using such an expansion, the Fourier moments of the response function from Eq.~\eqref{eq_moment} can be expressed as follows
\begin{equation}
m(j\tau)=\sum_{k=1}^{N_O}\sum_{l=1}^{N_O}o_ko_l m_{k,l}(j\tau)\;, 
\end{equation}
where we have introduced

\begin{equation}
\label{eq_expectation}
m_{k,l}(j\tau) = \langle0 \lvert (O_{k}B)^\dagger \mathcal{U}(j\tau) O_l B\rvert0 \rangle\;.
\end{equation}
In the above expression we identified $\mathcal{U}(j\tau)$ with the time evolution operator and $B$ with the initial state preparation unitary $B|0\rangle = |\Psi_0 \rangle$. These $N_O(N_O+1)/2$ expectation values can then be extracted from the quantum simulation. In situations where $N_O$ is large it might become beneficial to consider alternative strategies as described in Ref.~\cite{Roggero_exc_2020}. Such expectation values can be computed using a Hadamard test~\cite{Hadamard_test}, a particular case of quantum phases estimation. A Hadamard test uses the phase\hyp kickback mechanism to encode the targeted expectation value on an ancillary qubit by applying a controlled version of the observable.

The real (imaginary) part of the diagonal terms in Eq.~\eqref{eq_expectation} can be computed with the following circuit~\cite{QA_cleve}
\begin{equation}
\label{circ:had_test_diag}
\Qcircuit @C=1.2em @R=1em {
\lstick{|0\rangle}&\gate{H}&\gate{I (S^\dagger)}&\ctrl{1}&\gate{H}&\meter \\
	\lstick{|\psi_k\rangle}&{/^n}\qw&\qw&\gate{\mathcal{U}(j\tau)}&\qw&\qw \\
 }\,\,\,,
\end{equation}
assuming the initial state $|\psi_k \rangle =  O_k B\left|0 \right>$.
The corresponding expectation values can then be retrieved by measuring the ancilla in the computational basis. The number of samples required scales as $\mathcal{O}(1/\epsilon^2)$, where $\epsilon$ is the estimation error, and can be improved to $\mathcal{O}(1/\epsilon)$ using amplitude amplification techniques~\cite{amplitude_amplification}. For the off-diagonal terms with $k\neq l$ we can absorb the excitation operators within the unitary as $\tilde{\mathcal{U}}(j\tau) = O_k \mathcal{U}(j\tau) O^\dagger_l$. This translates into a circuit as 
\begin{equation}
\label{circ:had_test_off}
\Qcircuit @C=0.4em @R=0.5em {
\lstick{|0\rangle}&\gate{H}&\ctrl{1}&\ctrl{1}&\ctrl{1}&\gate{H}&\qw\\
	\lstick{|\Psi_0\rangle }&\qw&\gate{O_k}&\gate{\mathcal{U}(j\tau)}&\gate{O_l^\dagger}&\qw&\qw\\
 }\,\,\,,
\end{equation}
for the real part. 

In the following, we introduce two error mitigation techniques, echo verification and operator decoherence renormalization, and tailor them to computing expectation values via Hadamard tests. 

\subsubsection{Echo verification}
Echo verification~\cite{OBrien_verification} and dual state purification~\cite{PRA_DSV} are equivalent techniques aiming at mitigating noise by verifying if an error occurred during the circuit execution, discarding the components corresponding to these errors in final state and purifying it before computing expectation values. This can be achieved by un\hyp preparing the initial state, projecting on $|0\rangle^{\otimes n}$, effectively neglecting erroneous runs by post-selection, and finally rescaling expectation values to account for the probability of post-selection. More specifically, consider the above circuit Eq.~\ref{circ:had_test_off},
which produces the state 
\begin{equation}
\begin{split}
|\Phi \rangle &= \frac{1}{\sqrt{2}}\left( |\bar{0}\rangle \otimes |0\rangle +  B^\dagger O_k^\dagger \mathcal{U}(j\tau)O_l B |\bar{0}\rangle \otimes |1\rangle  \right) \\
&\equiv \frac{1}{\sqrt{2}}\left( |\bar{0}\rangle \otimes |\bar{0}\rangle +  \rvert\phi \rangle \otimes |1\rangle  \right),
\end{split}
\end{equation}
where $B$ is the ground state preparation circuit $|\Psi_0\rangle = B|0\rangle$ and $|\bar{0}\rangle= |0\rangle^{\otimes n}$. We remark that $|\phi \rangle$ can always be written in the basis spanned by $|\bar{0} \rangle$ and a state orthogonal to it as follows
\begin{equation}
 |\phi \rangle = \alpha |\bar{0}\rangle + \beta |\bar{0}^\perp \rangle.
\end{equation}
Calculation of the moment $m_{k,l}(j\tau)$ is then equivalent to the task to the estimation of the $\alpha$ coefficient, in fact
\begin{equation}
\alpha=\langle\bar{0}\vert\phi\rangle =\langle\bar{0}\lvert B^\dagger O_k^\dagger \mathcal{U}(j\tau)O_l B \rvert\bar{0}\rangle=m_{k,l}(j\tau)\;.
\end{equation}

In the basis composed by the normalized states $\{\rvert\bar{0}\rangle\otimes\rvert0\rangle,\rvert\bar{0}^\perp\rangle\otimes\rvert0\rangle,\rvert\bar{0}\rangle\otimes\rvert1\rangle,\rvert\bar{0}^\perp\rangle\otimes\rvert1\rangle\}$, the density matrix of the state $|\Phi\rangle$ can then be written as
\begin{equation}
\rho = \frac{1}{2}
    \begin{pmatrix}
        1 & 0 & \alpha^* & \beta^* \\ 
        0 & 0 & 0 & 0 \\
        \alpha & 0 & |\alpha|^2 & \alpha \beta^* \\
        \beta & 0 & \alpha^* \beta & |\beta|^2
    \end{pmatrix},
\end{equation}
which is reduced to the density matrix for the ancilla

\begin{equation}
\rho_0 = \Tr\left[\rho \left(\rvert\bar{0}\rangle \langle \bar{0}\lvert\otimes\mathbb{1}\right) \right]=\frac{1}{2}
    \begin{pmatrix}
        1 &  \alpha^*\\ 
        \alpha &  |\alpha|^2 \\
    \end{pmatrix}
\end{equation}
when projected onto $\rvert\bar{0}\rangle$ state of the main system.
The real and imaginary part of $\alpha$ can then be obtained from Pauli measurements on the ancilla qubit as
\begin{equation}
\begin{split}
\langle X_a& \rangle_0=\frac{2\Re{\alpha}}{1+|\alpha|^2},\quad\langle Y_a \rangle_0=\frac{2\Im{\alpha}}{1+|\alpha|^2}\\
\langle& Z_a \rangle_0=\frac{1-|\alpha|^2}{1+|\alpha|^2}=\frac{2}{1+|\alpha|^2}-1\;.
\end{split}
\end{equation}
In the above, $\langle \sigma_a \rangle_0$ denotes the expectation value of the Pauli matrix $\sigma$ on the ancilla post\hyp selected on the physical system being in the $\rvert\bar{0}\rangle$ state. These results can be retrieved by using $\langle \sigma_a \rangle_0 = \Tr{\sigma_a \rho_0}/\Tr{\rho_0}$.
As pointed out already in Ref.~\cite{OBrien_verification}, one can clearly see that if one uses only post-selection, the expectation values $\langle X_a \rangle_0$ and $\langle Y_a \rangle_0$ will give biased results. One can however use the expectation value $\langle Z_a \rangle_0$ to unbias the result~\cite{PRA_DSV}
\begin{equation}
    \begin{split}
        \Re{\alpha} &= \frac{\langle X_a \rangle_0}{1 + \langle Z_a \rangle_0},\quad
        \Im{\alpha} = \frac{\langle Y_a \rangle_0}{1 + \langle Z_a \rangle_0}.\\ 
    \end{split}
\end{equation}

The key observation is that projection onto the $\rvert\bar{0}\rangle$ state preserves the information we are attempting to extract, in our case the complex number $\alpha$, while at the same time removing part of the errors~\cite{OBrien_verification,PRA_DSV}. One can see this directly by considering the effect of depolarizing noise, described by the channel 
\begin{equation}
    \mathcal{N}(\rho) = (1-p)\rho + \frac{p}{2^n}\mathbb{1},
    \label{eq_depolarising_channel}
\end{equation}
on the expectation values. Indeed, the density matrix of the state $\rvert\Phi\rangle$ under a depolarising noise channel with parameter $p$ becomes 
\begin{equation}
\widetilde{\rho} = (1-p) \rho + \frac{p}{2^{n+1}}\mathbb{1}\;.
\end{equation}
Projection onto the state $\rvert\bar{0}\rangle$ leads to
\begin{equation}
\begin{split}
\widetilde{\rho}\left(\rvert\bar{0}\rangle \langle \bar{0}\lvert\otimes\mathbb{1}\right)=&(1-p)\rho\left(\rvert\bar{0}\rangle \langle \bar{0}\lvert\otimes\mathbb{1}\right)\\
&+\frac{p}{2^{n+1}}\rvert\bar{0}\rangle \langle \bar{0}\lvert\otimes\mathbb{1}\;,
\end{split}
\end{equation}
leading to a corresponding reduced density matrix on the ancilla qubit given by the following expression
\begin{equation}
    \widetilde{\rho}_0 =  (1-p)\rho_0 + \frac{p}{2^{n+1}} \mathbb{1}\;,
\end{equation}
where now the identity $\mathbb{1}$ is intended as a $2\times2$ matrix.
The noisy expectation values on the ancilla take now the following form
\begin{equation}
\begin{split}
\widetilde{\langle Z_a\rangle_0}=\frac {\Tr{Z_a \tilde{\rho}_0}}{\Tr{\tilde{\rho}_0}}%&=\frac{1-p}{2}(1-|\alpha|^2)\frac{1}{\frac{1-p}{2}(1+|\alpha|^2)+\frac{p}{2^n}}\\
%&=\frac{1-p+\frac{p}{2^n}-\frac{1-p}{2}(1+|\alpha|^2)-\frac{p}{2^n}}{\frac{1-p}{2}(1+|\alpha|^2)+\frac{p}{2^n}}\\
&=\frac{1-p+\frac{p}{2^n}}{\frac{1-p}{2}(1+|\alpha|^2)+\frac{p}{2^n}}-1\\
\end{split}
\end{equation}
\begin{equation}
\begin{split}
\widetilde{\langle X_a\rangle_0}=\frac {\Tr{X_a \tilde{\rho}_0}}{\Tr{\tilde{\rho}_0}}&=\frac{(1-p)\Re{\alpha}}{\frac{1-p}{2}(1+|\alpha|^2)+\frac{p}{2^n}}\\
\end{split}
\end{equation}
and analogously for $\widetilde{\langle Y_a\rangle_0}$. This leads to

\begin{equation}
\frac{\widetilde{\langle X_a\rangle_0}}{1+\widetilde{\langle Z_a\rangle_0}}=\Re{\alpha}+\mathcal{O}\left(\frac{p}{2^n}\right)\;,
\end{equation}
where now the bias has been reduced by a factor $\mathcal{O}(2^{-n})$ depending on the main system size~\cite{OBrien_verification}.

The only overhead of this technique is the increase in sample complexity due to the need to estimate the post-selected noisy expectation value with an error $P_0\epsilon$, with $P_0$ the probability of measuring the $\rvert\bar{0}\rangle$ state (see~\cite{OBrien_verification} for a more detailed discussion). This probability is given by
\begin{equation}
P_0=\frac{1-p}{2}(1+|\alpha|^2)+\frac{p}{2^n}>\frac{1-p}{2}\;,
\end{equation}
which can become small for a large noise level.
%ue to small probably of the physical system to end in the $|\bar{0}\rangle$ state. 
However, we empirically show that this probability remains above 0.08 for the experiments performed in this work.

In the noise\hyp free regime, the ancilla after post\hyp selection is in a pure state. Therefore, in the presence of noise, we can purify it to get closer to the ideal result \cite{PRA_DSV, OBrien-purification}. State tomography, see e.g. \cite{state-tomography-Gambetta}, on the ancilla qubit can be performed by measuring it in the $X$, $Y$ and $Z$ basis, which are anyway required in our scheme to obtain the real and imaginary parts. The expectation values are then computed on the closest pure state. We will call this method, proposed already in Ref.~\cite{OBrien-purification, PRA_DSV}, PEV. 

Even if we only discussed the robustness of PEV for depolarizing noise one can show that this strategy corrects also different noise models. For instance Ref.~\cite{PRA_DSV} shows that an orthogonal error channel is corrected from $p$ to $p^2/(M(1-p^2))$, where $p$ is the probability of ending into a subspace spanned by $M$ orthogonal states. On the other hand, Ref~\cite{OBrien_verification} provides numerical experiments for depolarising, amplitude and phase damping channel, demonstrating that the error decreases quadratically with $p$, in the case of a Givens rotation circuit. 

To summarise, purified echo verification is a powerful technique to mitigate errors for computing expectation values. It has been shown to be robust against various error channel. In the remaining of this paper, we will demonstrate that it is also effective on real quantum hardware. 

\subsubsection{Operator decoherence renormalization}

The operator decoherence renormalization strategy \cite{mitigating_depolarising, farrell2023scalable}, also referred to as self\hyp mitigation \cite{self-mitigation}, estimates the parameters of the assumed noise model and used it to revert the effect of the noise. 
The expectation value of a Pauli observable under a depolarising noise channel, see Eq.~\eqref{eq_depolarising_channel}, is 
\begin{equation}
    \Tr{\sigma \mathcal{N}(\rho)} = (1-p) \Tr{\sigma \rho},
\end{equation}
which can be corrected by dividing by $(1-p)$. The main idea is then to run a circuit with known expectation value and infer the noise parameter. For instance, since a backward step cancels a forward step, we have 
\begin{equation}
    1 = \langle \bar{0} | BO_l \mathcal{U}(j\tau)\mathcal{U}(-j\tau)O^\dagger_l B^\dagger |\bar{0} \rangle,
\end{equation}
which becomes $(1-p)$ under the depolarising channel. 

Thus, we can correct an even number of Trotter steps by running an additional noise\hyp estimating circuit, with a reference initial state $|\psi\rangle$, 
\begin{equation}
\label{circ:ODR}
\Qcircuit @C=1.2em @R=1em {
\lstick{|0\rangle}&\gate{H}&\ctrl{1}&\ctrl{1}&\gate{H}&\meter \\
	\lstick{|\psi \rangle}&\qw&\gate{\mathcal{U}(j\tau)}&\gate{\mathcal{U}(-j\tau)}&\qw&\qw \\
 }\,\,\,,
\end{equation}
as
\begin{equation}
\label{eq:sv}
\langle \psi| \TE{2j\tau}|\psi_k\rangle_{ODR} = \frac{\langle \psi |\TE{2j\tau} |\psi\rangle}{\langle\Psi| \TE{j\tau}\TE{-j\tau}|\psi \rangle}.
\end{equation}
We note that the noise renormalization circuit can only be run for the diagonal $l=k$ part. However, if the dominant contribution to the error comes from the time evolution, and not the excitation operator, it is still reasonable to use it to correct the off\hyp diagonal ones.

Since this scheme assumes depolarising noise, we might wonder how it can still be useful on a real device, where the noise is more complicated. In fact, it is important to make the noise look more depolarising, using techniques such as Pauli twirling and randomized compiling \cite{randomOuyang2020compilation,Randomized_compiling,Pauli_Twirling}. 

\subsection{Control Reversal Gates}
\label{CRG}
Even if the CNOT overhead from the control operations only scales linearly with the number of terms in the Hamiltonian, it is valuable, and particularly when working with real devices, to reduce this cost as much as possible. While ancilla\hyp free techniques do exist, see e.g. \cite{Lu_2021,OBrien_verification,QPE_ancilla_free}, we instead choose to reduce the overhead by using control reversal gates (CRG), see e.g. \cite{QETU,CRG}. A reversal gate $R$ is a product of Pauli matrices which anti\hyp commutes with the Hamiltonian, i.e., $\{H,R\}=0$. The control reversal gate consists of its controlled version on an ancilla being $|0\rangle$. CRG enable toggling the flow of the time evolution in a forward or backward manner. For instance, if the ancilla is in the zero state, we apply 
\begin{equation}
\begin{split}
    &R \exp{-iHt}R^\dagger = R\sum_n \frac{(-itH)^n}{n!}R^\dagger =\\
    &\sum_n \frac{(itH)^n}{n!}RR^\dagger = \exp{itH},
    \end{split}
\end{equation}
which is a backward time evolution, and simply a forward step $\exp{-iHt}$ otherwise. The CRG framework has the additional advantage of needing only half time simulations, since the phase kickback contribution happens twice. Thus, only half of the number of Trotter steps are required to reach the same accuracy, up to a factor of two. The real part of the diagonal moment $m_{k,k}(j\tau)$ can be evaluated using the following circuit (cf. Eq.~\eqref{circ:had_test_diag})
\begin{equation}
\label{circ:CRG}
\Qcircuit @C=1em @R=1em {
\lstick{|0\rangle}&\gate{H}&\ctrlo{1}&\qw&\ctrlo{1}&\gate{H}&\meter \\
	\lstick{|\psi_k\rangle}&{/^n}\qw&\gate{R}&\gate{\mathcal{U}(j\tau/2)}&\gate{R^\dagger}&\qw&\qw \\
 }\,\,\,.
\end{equation}
It is important to note that the backward time evolution must be the inverse of the forward time evolution, i.e., $\mathcal{U}(-t)\mathcal{U}(t)=\mathbb{1}$, and hence, $\mathcal{U}(-t)=\mathcal{U}(t)^{-1}$. This is automatically satisfied for product formulas of even order. 

It might not always be possible to find a single CRG for the full Hamiltonian, as in Eq.~\eqref{circ:CRG}. In that case, $H$ can be split into $M$ groups $H=\sum_m^M H_m$, and individual CRG can be found for each of them. The CRGs can now be inserted between each Trotter steps as~\cite{QETU} 
\begin{equation}
\prod_i^r \prod_m^M R_m \mathcal{U}_1(t/r)R_m^\dagger,
\end{equation}
where only $R_m$ are anti-controlled on the ancilla. In the worst\hyp case scenario we can show that it suffices to take
\begin{equation} 
\label{eq:num_cr}
M = 2n,
\end{equation} 
where $n$ is the number of qubits. The grouping can be achieved as follows: for every term in the Hamiltonian we start grouping together all terms that have an $X$ or $Y$ operation at location $0$ and in a separate group all terms with a $Z$ operation at location $0$, we can then choose $R=Z_0$ for the first group and $R=X_0$ for the second. The procedure is then repeated for all $n$ location obtaining $M=2n$ groups that can be reversed using single qubit Pauli operators. Using this strategy and for a general $n$ qubit Hamiltonian, a single first order Trotter step controlled on an ancilla qubit can be implemented with an additive cost of $4n$ CNOT gates and $2n$ Hadamard gates over the base cost of implementing a first the Trotter step without a control. In comparison, a naive implementation of the controlled first order Trotter step for an Hamiltonian $H$ composed of $\Gamma$ Pauli strings, where each of them is exponentiated individually, requires $2\Gamma$ CNOT gates.
In the case $\Gamma>2n$, CRG provide an advantage over the direct implementation. In particular, the overhead scales always at most linearly in the system size whereas a direct implementation for a $k$-local Hamiltonian will require $\mathcal{O}(n^k)$ additional entangling gates. To the best of our knowledge this is a novel result which might be of interest in other applications besides the estimation of Fourier moments.

We also note that the estimate in Eq.~\eqref{eq:num_cr} is for the worst\hyp case scenario and that much fewer groups can be found on a case\hyp by\hyp case basis. For instance, Heisenberg models with arbitrary coefficients on one dimensional spin chains only requires two groups with corresponding reversal gates given by~\cite{QETU}
\begin{equation}
\begin{split}
R_Z&=\bigotimes_{i=0}^{\lfloor n/2 \rfloor } Z_{2i}\quad\quad R_X=\bigotimes_{i=0}^{\lfloor n/2 \rfloor} X_{2i}\;.
\end{split}
\end{equation}

For the Hamiltonian considered in this work we only require $M=3$. When implementing $r$ second order Trotter steps however, one can exploit cancellations to reduce the number of CRG to a total of $2(1+r)$. With linear connectivity this can be implemented using $14+2r$ CNOT gates in total, on top of the gate requirements for the un\hyp controlled. The derivations are covered in more details in App.~\ref{explicit circuit}.

Alternatively, suppose we have access to multi\hyp qubit gates, such as the M\o lmer\hyp S\o rensen gate~\cite{MS} in trapped ion devices. In that case, it is advantageous to consider the effective Hamiltonian $\tilde{H} = Z\otimes H$, with CRG given by $R=X\otimes \mathbb{1}_n$ and initial state $|0\rangle \otimes |\Psi\rangle$. In this configuration, and assuming that simulating $\tilde{H}$ has the same cost as simulating $H$, the Hadamard test can be performed in an almost control\hyp free manner. 

\subsubsection{Echo verification}
The echo verification procedure described in the context of the standard Hadamard test above can be easily transferred to control reversal gates by considering an even number of Trotter steps. For instance, if we consider two Trotter steps we have the following identities 
\begin{equation}
\begin{split}
&\Qcircuit @C=0.4em @R=0.5em {
&\gate{H}&\ctrl{1}&\ctrl{1}&\ctrl{1}&\qw &\qw \\
&\gate{B}\qw&\gate{O_k}&\gate{\mathcal{U}(j\tau)^2}&\gate{O_l^\dagger}&\gate{B^\dagger}&\qw\\
 }= \\
 & \Qcircuit @C=0.2em @R=0.8em @!R {
&\gate{H}&\ctrl{1}&\qw&\ctrl{1}&\ctrlo{1}&\ctrl{1}&\qw \\
&\gate{B}&\gate{O_k}&\gate{\mathcal{U}(j\tau)}&\gate{\mathcal{U}(j\tau)}&\gate{\mathcal{U}(j\tau)^\dagger}&\gate{O_l^\dagger}&\gate{B^\dagger}&\qw 
 } =\\
 & \Qcircuit @C=0.2em @R=0.8em @!R {
&\gate{H}&\ctrl{1}&\qw&\ctrlo{1}&\qw&\ctrlo{1}&\ctrl{1}&\qw \\
&\gate{B}&\gate{O_k}&\gate{\mathcal{U}(j\tau)}&\gate{R}&\gate{\mathcal{U}(j\tau)}&\gate{R^\dagger}&\gate{O_l^\dagger}&\gate{B^\dagger}&\qw 
 }\,\,\,.
 \end{split}
 \end{equation}

In the first step, we use an anti-controlled backward evolution to cancel the first evolution when the ancilla is in the zero state, while in the second step we use the CRG scheme to replace the two controlled evolutions. 
Hence, circuits with CRG can employed within the framework used in the EV protocol. However, note that the fast forwarding factor from CRG is lost since we use the backward propagation to cancel the forward contribution when the ancilla is in the $|0\rangle$ state. 
\subsubsection{Operator decoherence renormalization}
Unlike EV, the ODR scheme can be directly used in conjunction with CRG by estimating the noise factor as before. For the evaluation of off\hyp diagonal moments $m_{k,l}(j\tau)$ with $k\neq l$ we need however to modify the base circuit from Eq.~\eqref{circ:ODR}. In order to do this, right after preparing the initial state with $B$ unitary we preform $A_i$ unitary operations controlled on the state ($i=0$ or $i=1$) and follow them with the CRG Hamiltonian evolution for a time $\tau/2$. The final state of the full system reads
\begin{equation}
    \begin{split}
    \TE{-j\tau/2} \frac{A_0}{\sqrt{2}}\rvert\Psi_0\rangle \otimes |0\rangle&+\TE{j\tau/2}\frac{A_1}{\sqrt{2}}\rvert\Psi_0\rangle\otimes |1\rangle\;,
    \end{split}
    \end{equation}
where the ancilla is starting in $\rvert+\rangle$ at the beginning. By applying an Hadamard gate to the ancilla we then find
    \begin{equation}
        \begin{split}
    &\frac{1}{2} \left(\TE{-j\tau/2}A_0+\TE{t/2}A_1\right)\rvert\Psi_0\rangle\otimes |0\rangle \\
    +& \frac{1}{2} \left(\TE{-j\tau/2}A_0-\TE{j\tau/2}A_1\right)\rvert\Psi_0\rangle\otimes |1\rangle .
    \end{split}
    \end{equation}
   The Pauli Z expectation value of the ancilla is then
    \begin{equation}
    \begin{split}
    \langle Z \rangle_a =&\frac{1}{4}\langle\Psi_0\lvert\big(2\mathbb{1}+A^\dagger_0\TE{j\tau/2}\TE{j\tau/2}A_1 \\
    +&A^\dagger_1\TE{-j\tau/2} \TE{-j\tau/2}A_0\big)\rvert\Psi_0\rangle\\
    -&\frac{1}{4}\langle\Psi_0\lvert\big(2\mathbb{1}-A^\dagger_0\TE{j\tau/2}\TE{j\tau/2}A_1\\
    -&A^\dagger_1\TE{-j\tau/2}\TE{-j\tau/2}A_0\big)\rvert\Psi_0\rangle\\
    =&\Re\left(\langle\Psi_0\lvert A^\dagger_0\TE{j\tau/2}\TE{j\tau/2}A_1\rvert\Psi_0\rangle\right)\;.
    \end{split}
    \end{equation}
    To obtain the desired value for the $m_{k,l}(j\tau)$ moment, we can then choose the unitaries as $A_0=O_l^\dagger$ and $A_1=O_k$. The imaginary component can be computed in a similar way adding an $S^\dagger$ gate to the ancilla.

\section{Physical model}
\label{physical_model}
\subsection{Nuclear lattice model}
We consider a model inspired by a pionless lattice effective field theory~\cite{eft}, and in particular, the simple toy model for a triton introduced in~\cite{neutrio_nucleus_roggero} and further studied in~\cite{two_point_roggero}. We consider $A= 2$ dynamical nucleons, together with a static one (infinite mass) fixed on the first site of a $2d$ lattice of size $L\times L$ with periodic boundary conditions. Even if this model is quite simple and can be easily simulated, it yet contains much of the leading order contributions to the interaction and can thus provide valuable information about light nuclei and their response function. The Hamiltonian of this model is formally equivalent to a $2d$ Fermi Hubbard model with hopping term
\begin{equation}
H_{\text{kin}}=-t \sum_{f=\{\uparrow,\downarrow\}}\sum_{\langle i,j\rangle}c^\dagger_{i,f}c_{j,f}\;,
\end{equation}
a two-body contact interaction%together with two and three\hyp body interactions
\begin{equation}
    \begin{split}
        H_{\text{int}} = &U \sum_{i=1}n_{i,\uparrow}n_{i,\downarrow}
%        H_{\text{int}} = &U \sum_{i=1}\sum_{f,f'}^{N_f}n_{i,f}n_{i,f'} + V \sum_{i=1}\sum_{f<f'<f''}n_{i,f}n_{i,f'}n_{i,f''}
    \end{split}
\end{equation}
and additional one and two-body potentials generated by the static proton at lattice site $i=1$
%as well as the potential created by the static nucleon 
\begin{equation}
 H_{\text{static}} = U \sum_{f=\{\uparrow,\downarrow\}}n_{1,f}+Vn_{1,\uparrow}n_{1,\downarrow}\;.
 \end{equation}
 We recall that the fermionic operator $c_{i,f}$ destroys a particle of the species $f$ on site $i$, $c^\dagger_{i,f}$ is the corresponding creation operator, and $n_{i,f} = c^{\dagger}_{i,f}c_{i,f}$ the number operator. The kinetic term contains a sum over $\langle i,j\rangle$ neighboring sites. The coefficient $U$ and $V$ correspond to the two\hyp body and three-body interaction strengths respectively. Realistic numerical values for $t,\,U$ and $V$ with a physical lattice spacing $a=1.4$fm can be found in \cite[Table 1]{neutrio_nucleus_roggero} (taken from \cite{Rokash_2014}). We use first quantization to encode the Hamiltonian into $2 \lceil \log(L^2) \rceil$ qubits, where $L$ is the number of sites per dimension.%, by mapping each site $|i\rangle$ to $|\text{bin}(i)\rangle$, where bin($x$) is the binary representation of $x$. 
Using a Gray code ordering of states of the lattice helps in reducing the complexity of the hopping term~\cite{neutrio_nucleus_roggero,DiMatteo_2021}. For a small system with $L=2$ this corresponds to using the following mapping~\cite{neutrio_nucleus_roggero}
\begin{equation}
    |1\rangle \equiv |00\rangle ~~ |2\rangle \equiv |01\rangle ~~ |3\rangle \equiv |10\rangle ~~ |4\rangle \equiv |11\rangle\;.
\end{equation}
As in Refs.~\cite{neutrio_nucleus_roggero,two_point_roggero} we consider a simplified setting with $t=1,\, U = -7$ and $V=-4U$, where most of the terms in the Hamiltonian are canceled,
the Hamiltonian for the model can be expressed in the Pauli basis as
\begin{equation}
\begin{split}
H &= 4.5\cdot\mathbb{1} -2 \sum_{i=1}^4 X_i \\
&+1.75 \left(\sum_{i<j<k} Z_iZ_jZ_k + Z_1Z_4 + Z_2Z_3 \right) .
\end{split}
\label{eq:nuclattice}
\end{equation} 
Here, $X_k,\,Y_k,\,Z_k$ are the corresponding Pauli matrices acting on qubit $k$. More details about the implementation of Hamiltonian evolution for pion-less EFT interactions can be found in Ref.~\cite{neutrio_nucleus_roggero} and Ref.~\cite{watson2023quantum}. Without loss of generality, we shift the Hamiltonian to cancel the terms proportional to the identity, i.e., by considering 
\begin{equation}
    \widetilde{H} = H - 4.5\cdot\mathbb{1}.
\end{equation}
This allow us to present the results in a more straightforward fashion. 

\subsection{Excitation operators}
In this preliminary study, similarly to what was done in Ref.~\cite{two_point_roggero}, we limit our discussion to scattering experiment where the probe couples to the nucleon density by transferring a momentum $\vec{q}$ and an energy $\omega$ to the target. Owing to the use of a periodic spatial lattice, the allowed momenta are quantized as

\begin{equation}
    \vec{q}_k = \frac{\pi}{L a}\vec{x}_k,
\end{equation}
where $L a$ is the spatial length of the lattice and $\vec{x}_k \in \mathbb{N}^2$ the position of the $k$\hyp th momentum on the reciprocal lattice. The excitation operator takes then the form%we focus on probes coupling to the nucleon density $n_{i,f}$, described by the following operator
\begin{equation}
\hat{O}(\vec{q}_k) = \sum_{f={\uparrow,\downarrow}}\rho_f(\vec{q}_k) = \sum_{f={\uparrow,\downarrow}} e_f\sum_i e^{i \vec{q}_k\cdot \vec{r}_i} n_{i,f},
\end{equation}
where $e_f$ denotes the charge of the nucleon $f$, $\rho_f(\vec{q}_k)$ the nucleon densities in momentum space and $\vec{r}_i$ the position of site $i$ on the spatial lattice. The results shown in the next sections are obtained for only one possible momentum transfer equal to $\vec{q}=\pi/(La)(0,1)$ and using unit charges $e_\uparrow=e_\downarrow=1$. In the Pauli basis the excitation operator used in this work takes the form (cf.~\cite{two_point_roggero})%Without loss of generality, the momentum transfer is fixed in the first site for the remainder of this paper. Thus the momentum transfer operator can we written  as 
\begin{equation}
    \Obs = Z_1 + Z_3\;.
\end{equation}

\subsection{Variational ground state preparation}
\label{vqe_sec}
The Variational Quantum Eigensolver (VQE) \cite{vqe_original} is an algorithm to prepare approximate ground states. The VQE minimizes the energy expectation value of a parameterized wave function, i.e., an ansatz, which should be close to the true ground state upon the convergence of the optimization procedure. Even if it is challenging to have theoretical guarantees on the accuracy of the VQE scheme, it has been empirically shown to be successful in areas such as quantum chemistry \cite{VQE_Gambetta,UCC-chemistry}, in frustrated magnetic systems \cite{Monaco_PRB,lmg_grossi}, or in nuclear physics \cite{Papenbrock-Deuterium, Stetcu2022,nuclear_shell_VQE, PRC_Kiss}.

We use the following ansatz already used in Ref.~\cite{neutrio_nucleus_roggero} 
\begin{equation}
\Qcircuit @C=1em @R=1em { 
	&\qw&\gate{Ry(\theta_0)} &\ctrl{2}&\qw&\qw&\ctrl{2}&\qw &\qw\\
    &\qw&\gate{Ry(\theta_0)} &\qw&\ctrl{1}&\qw&\qw&\ctrl{1}&\qw\\
       &\qw&\gate{Ry(\theta_0)} &\qw&\ctrl{0}&\gate{Ry(\theta_1)}&\qw&\ctrl{0}&\qw\\
       &\qw&\gate{Ry(\theta_0)} &\ctrl{-1} &\qw&\gate{Ry(\theta_1)}&\ctrl{-1}&\qw&\qw\\
       },
\end{equation}
which is parameterized by two angles $\theta_0$ and $\theta_1$ encoded via rotation around the $y-$axis $Ry(\theta) = \exp{-i\theta Y}$, and is implementable with only four CZ gates with linear connectivity. 
The UCCSD entanglement structure inspires the form of the quantum circuit we would expect in the $V=0$ regime (without three\hyp body interactions), and by the fact that the Hamiltonian is real in the computational basis. Despite its simplicity, the ansatz achieves less than 10\% relative error of the exact ground state energy. The parameters are optimized using the gradient\hyp free optimizer COBYLA \cite{cobyla} on a simulator and are then fixed for the remaining of the work. %Extensive experimentation suggests that this ansatz is the most accurate strategy using a quantum circuit that can be run on current quantum hardware, probably due to the model's simplicity. 

\section{Results}
\label{res_sec}
In this section, we present the computations of the first few moments using the noise mitigation strategy presented above. We present an experiment with a a double Trotter step and a second one with multiple double steps aiming at reaching the breaking point of the methods. With the term "double Trotter step" we refer to two Trotter steps with half time each, $\mathcal{U}(j\tau) =\mathcal{U}(j\tau/2)\mathcal{U}(j\tau/2)$, since we are limited to perform an even number of steps by our error mitigation scheme. 

\subsection{Noise model}
\begin{figure}
    \centering
    \includegraphics[scale=0.45]{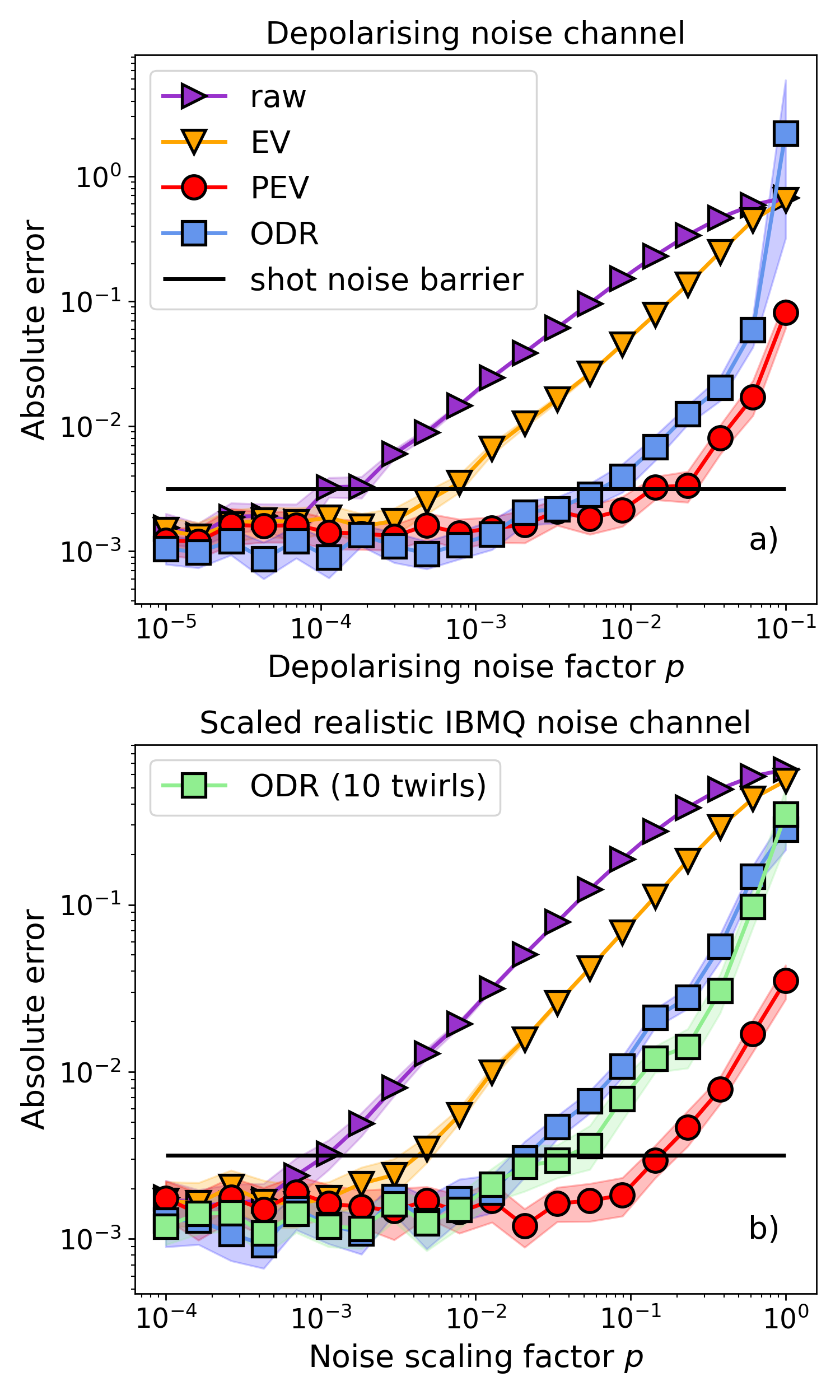}
    \caption{\textbf{Noise model:} Absolute error against the strength $p$ of the noise model for different error mitigation strategy. A depolarising noise model is used in the upper panel a), while in the bottom b) we have a scaled fake backend from qiskit, with noise parameters $p\in [p_0 \times 10^{-4},p_0]$, $p_0$ being the realistic coefficient. Note that in the first case, the circuit consist of a double Trotter step with 56 CNOT gates while in the second we have 12 double steps composed of 452 CNOT gates. The black line shows an upper bound on the  shot noise barrier. The shaded area correspond to a 95\% confidence interval obtained over 30 repetitions.}
    \label{fig:noise}
\end{figure}

Before going to the real quantum computer, we test (P)EV and ODR on two noise models with varying strength. Fig.~\ref{fig:noise} displays the absolute error against the strength of the noise model in the case of a depolarising channel in panel a) and a scaled fake backend from qiskit in panel b). In the first case, the circuit consist of a single double Trotter step with 56 CNOT gates and $O_i=O_j=Z_1$, while in the second we have 12 double steps composed of 452 CNOT gates.

We observe that the purified EV is the best performing technique, and crosses the shot noise barrier, i.e. the bound on the statistical error with $10^5$ shots, at a value of $p$ one hundred times greater than for the raw data. In panel b), we can see a marginal improvement for the ODR strategy due to Pauli twirling. Note that twirling has no effect when the noise is depolarising and only a negligible one on the (P)EV.

\subsection{Single step on hardware}
We now present results obtained on the superconducting quantum device ibmq\_kolkata~\cite{Jurcevic_2021}, containing 27\hyp fixed\hyp frequency transmon qubits, with fundamental transition frequencies of approximately 5 GHz and anharmonicities of $-340$ MHz. Microwave pulses are used for single\hyp qubit gates and cross\hyp resonance interactions~\cite{cross-resonance} are used for two\hyp qubit gates. The median qubit lifetime $T_1$ of the qubits is 109.86 $\mu s$ and the median coherence time $T_2$ is 58.95 $\mu s$. The qubits used in the experiments are chosen by hand and have a CNOT error varying between $6\times 10^{-3}$ and $9\times 10^{-3}$, readout error between $7\times 10^{-3}$ and $4\times 10^{-2}$ and sx error between $3\times 10^{-4}$ and $4\times 10^{-4}$. 

In the first experiment, we compute the ten first moments using a double second\hyp order Trotter step with $\tau = 0.125$. Error suppression and calibration techniques such as XY-8 dynamical decoupling \cite{DD_Lloyd,robust_DD,sequence_DD,sequence_DD2, genetic_DD}, pulse efficient transpilation \cite{pulse-efficient}, Pauli twirling \cite{Pauli_Twirling,Randomized_compiling} with 16 samples, control mitigation \cite{OBrien_verification} and readout error mitigation \cite{Nation_PRXQ} are always included, except explicitly stated otherwise. Every circuit is run with $10^5$ shots. Details about the implementation can be found in App.~\ref{error_suppresion}. Even if twirling is not strictly necessary when using PEV, we use it by default since it remains standard practice and does increase the sampling overhead, as the shot budget is distributed among the twirls. However, it remains necessary for ODR to work efficiently. 

\begin{figure}
    \centering
    \includegraphics[scale=0.38]{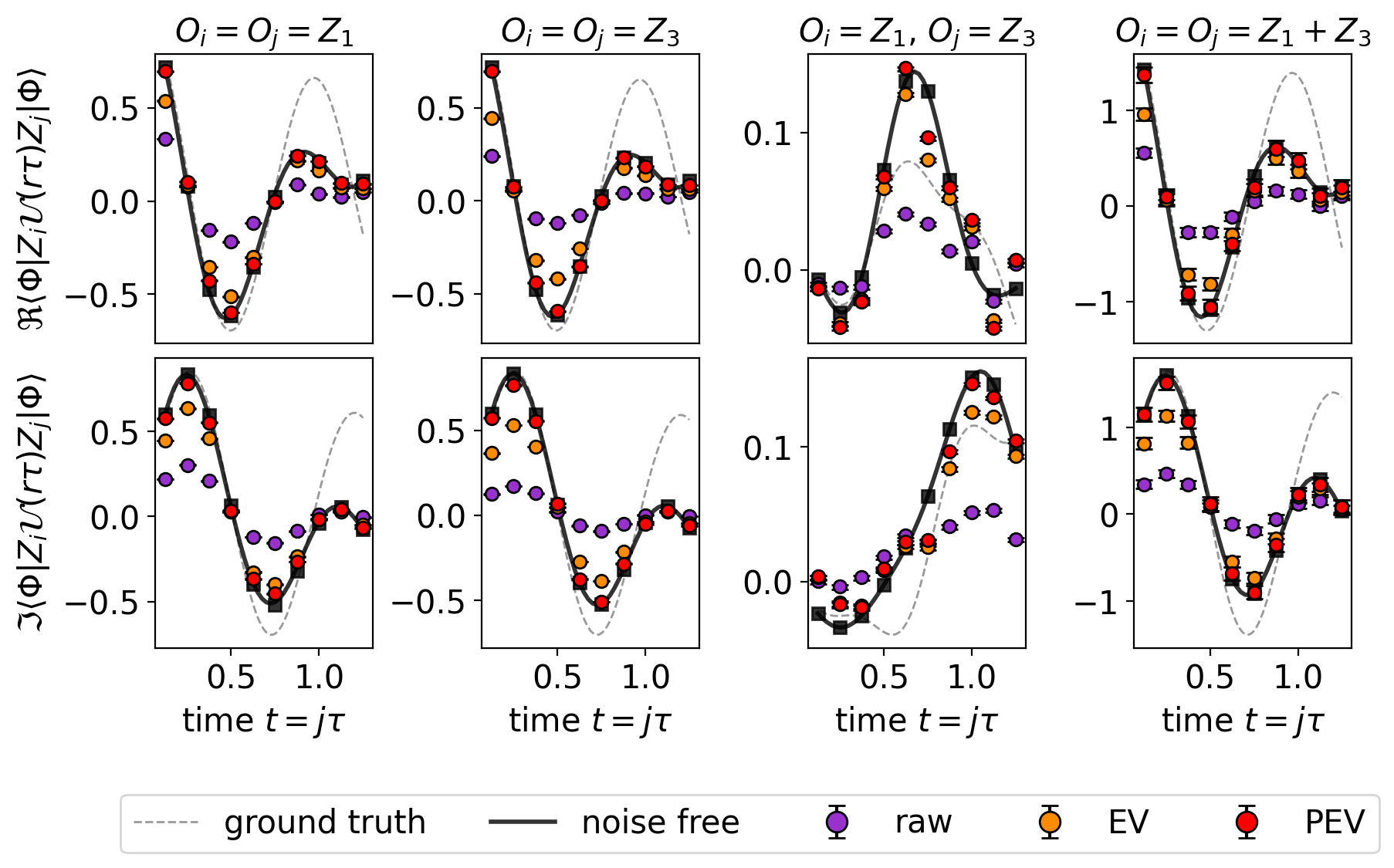}
    \caption{\textbf{Echo verification - single step:}  The first two columns display the diagonal components, the third one the off\hyp diagonal one, and in the last column we find the reconstructed moments. The real and imaginary part are displayed in the first and second row respectively. The grey curve indicates the ground truth from exact matrix multiplication, the black dotts are the noise\hyp less simulation while the colored ones are obtained on ibmq\_kolkata with EV (orange) and PEV (red). The errorbars correspond to one standard deviation computed via Bayesian inference.} 
    \label{fig:EV}
\end{figure}
Fig.~\ref{fig:EV} shows the results using echo verification. The first two columns display the diagonal components, the third one the off\hyp diagonal one, and in the last column we show the entire reconstructed moments. The real and imaginary part are displayed in the first and second row respectively. The errorbars correspond to one standard deviation computed via Bayesian inference~\cite{PRD_neutrino}, see App.~\ref{bayesian} for more details. We observe that the data obtained with PEV match the noise\hyp less simulation up to a small error, while the raw data are damped. Secondly, the Trotter approximation break downs after a time of $t=0.5$, indicating that more Trotter steps are required. 

\begin{figure}
    \centering
    \includegraphics[scale=0.38]{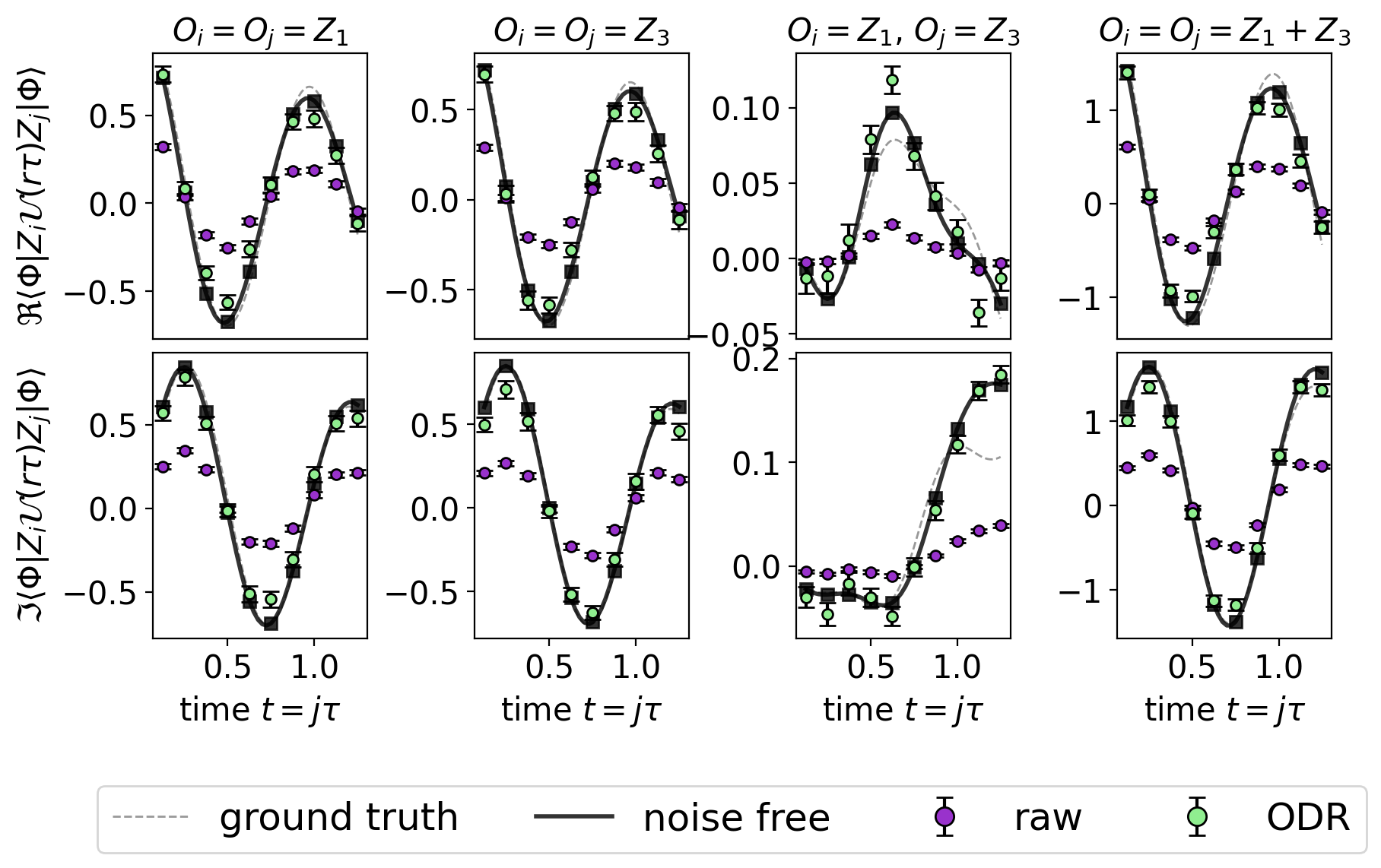}
    \caption{\textbf{Operator decoherence renormalization - single step:} The first two columns display the diagonal component, the third one the off\hyp diagonal one and the last the full moment. The real and imaginary part are displayed in the first and second row respectively. The grey curve indicates the ground truth from exact matrix multiplication, the black dotes are the noise less simulation while the colored ones are obtained on ibmq\_kolkata with ODR (green). The errorbars correspond to one standard deviation computed via Bayesian inference. } 
    \label{fig:NR}
\end{figure}

Fig.~\ref{fig:NR} displays the same quantities but using instead the operator decoherence renormalization strategy. ODR seems to be as effective as PEV in this case, even if the off\hyp diagonal terms have a larger variance. This can be explain by the fact that $(1-p)$ is estimated on the diagonal elements instead, and also by their lower scale. Even if the effect of Pauli twirling is not dramatic, it appears to reduce the variance. Finally, this time the Trotter approximation remains accurate for longer times, which is because we benefit from the factor of two fast forwarding provided by the CRG, which is an important benefit of this strategy.

We asses the statistical compatibility of the noise-mitigated moments $m$ with the noise\hyp free expected values $\mu$ by the mean of the $z$ statistical test. The average errors squared over the empirical variance $\sigma^2$
\begin{equation}
z^2 = \frac{1}{N} \sum_{i=1}^{N} \frac{(m_i - \mu_i)^2}{\sigma_i^2}
\end{equation}
are reported in Table~\ref{tab:my_label}. In the case of PEV, values are close to unity. We conclude that the data falls within one standard deviation around the expected values. For the ODR protocol, data falls within three standard deviation. Overall, the two error mitigation strategies provide an improvement of around one order of magnitude compared to the raw data. 

\begin{table}
    \centering
    \begin{tabular}{c|cc||cc}
        $z$ &raw&PEV&raw & ODR \\
         \hline
      real  &10.47 &0.62& 23.87&2.27  \\
      imag  &13.23 &0.63 & 31.97&2.03 \\
    \end{tabular}
    \caption{\textbf{Statistical analysis:} The z-score is reported for the real and imaginary part of the final moments computed with the purified PEV and ODR noise mitigation protocols.}
    \label{tab:my_label}
\end{table}

\subsection{Multiple steps on hardware}
\begin{figure}
    \centering
    \includegraphics[scale=0.7]{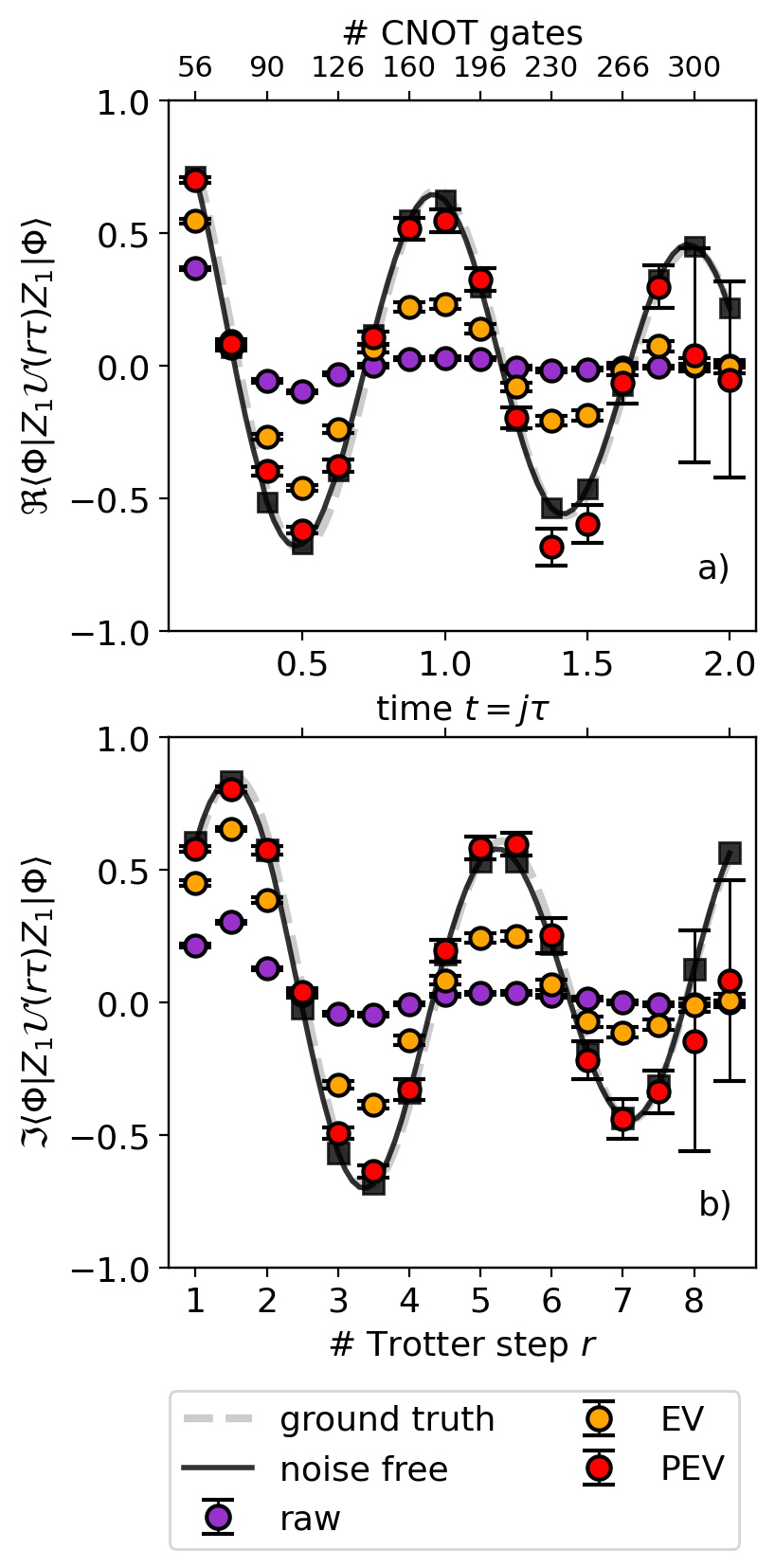}
    \caption{\textbf{Echo verification - multiple steps:}  The upper (lower) panel shows the real (imaginary) part of the first diagonal contribution to the moment, with $O_l=O_k=O_1 = Z_1$. The grey curve indicates the ground truth from exact matrix multiplication, the black dotes are the noise less simulation while the colored ones are obtained on ibmq\_kolkata with EV (orange) and PEV (red). The errorbars correspond to one sigma computed via Bayesian inference. The number of CNOT gates are indicated in the top axis, while the time is indicated between the two panels, and number of Trotter steps at the bottom.}
    \label{fig:step_EV}
\end{figure}

\begin{figure}
    \centering
    \includegraphics[scale=0.7]{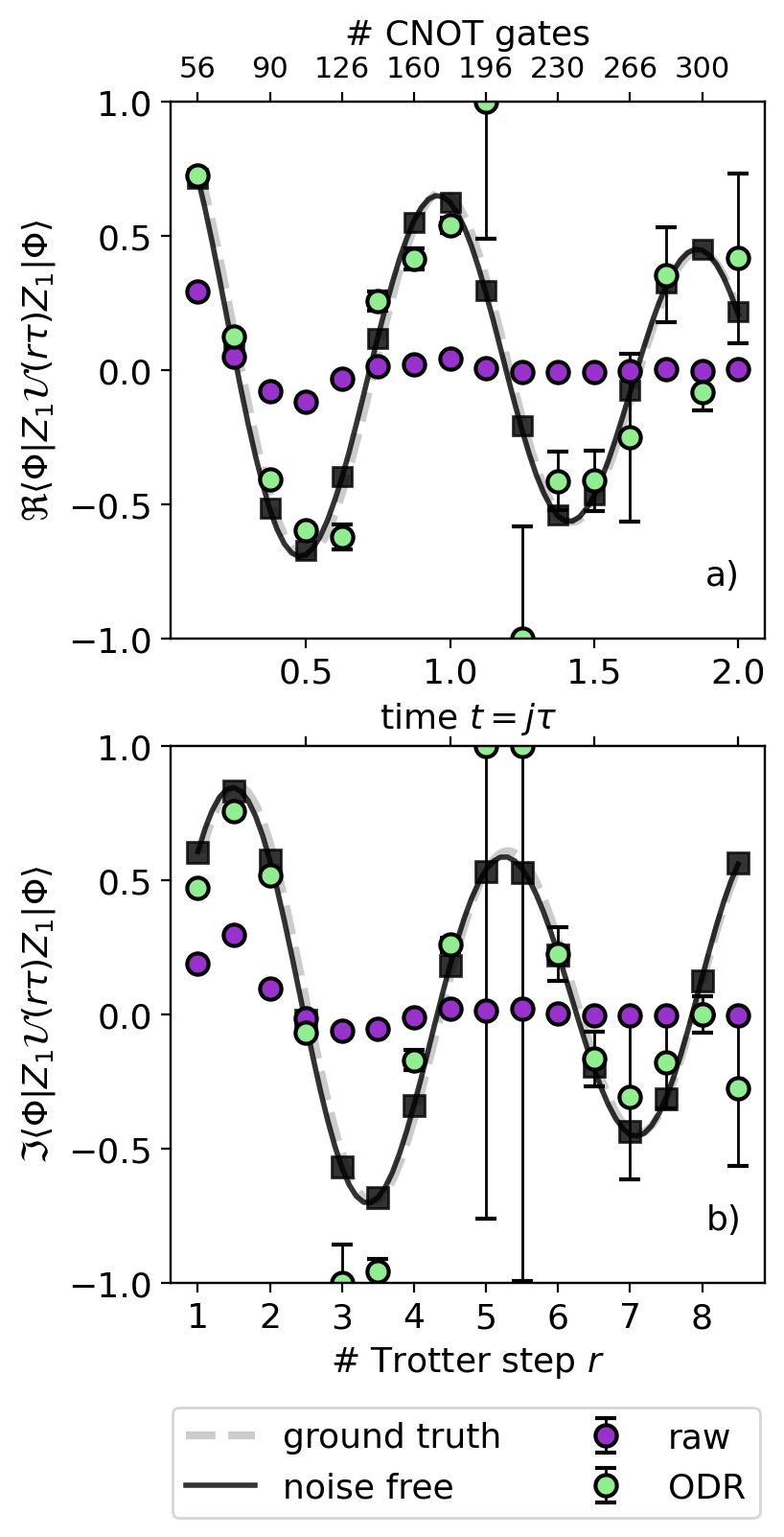}
    \caption{\textbf{Operator decoherence renormalization - multiple steps:}  The upper (lower) panel shows the real (imaginary) part of the first diagonal contribution to the moment, with $O_l=O_k=O_1 = Z_1$. The grey curve indicates the ground truth from exact matrix multiplication, the black dotes are the noise less simulation while the colored ones are obtained on ibmq\_kolkata with ODR (green). The data that are out of scale, due to a too small renormalization factor, are set to $\pm 1$. The errorbars correspond to one sigma computed via Bayesian inference. The number of CNOT gates are indicated in the top axis, while the time is indicated between the two panels, and number of Trotter steps at the bottom.}
    \label{fig:step_NR}
\end{figure}

In our second experiment, we increase the number of Trotter steps up to the breaking point. For simplicity, we only compute the real and imaginary part of the first diagonal contribution to the moment, with $O_l=O_k=O_1 = Z_1$, which are displayed in Fig.~\ref{fig:step_EV} and in Fig.~\ref{fig:step_NR} for the PEV and ODR mitigation strategy respectively. We perform up to eight double Trotter steps, summing up to a total of 300 CNOT gates. 

\begin{figure}
    \centering
    \includegraphics[scale=0.6]{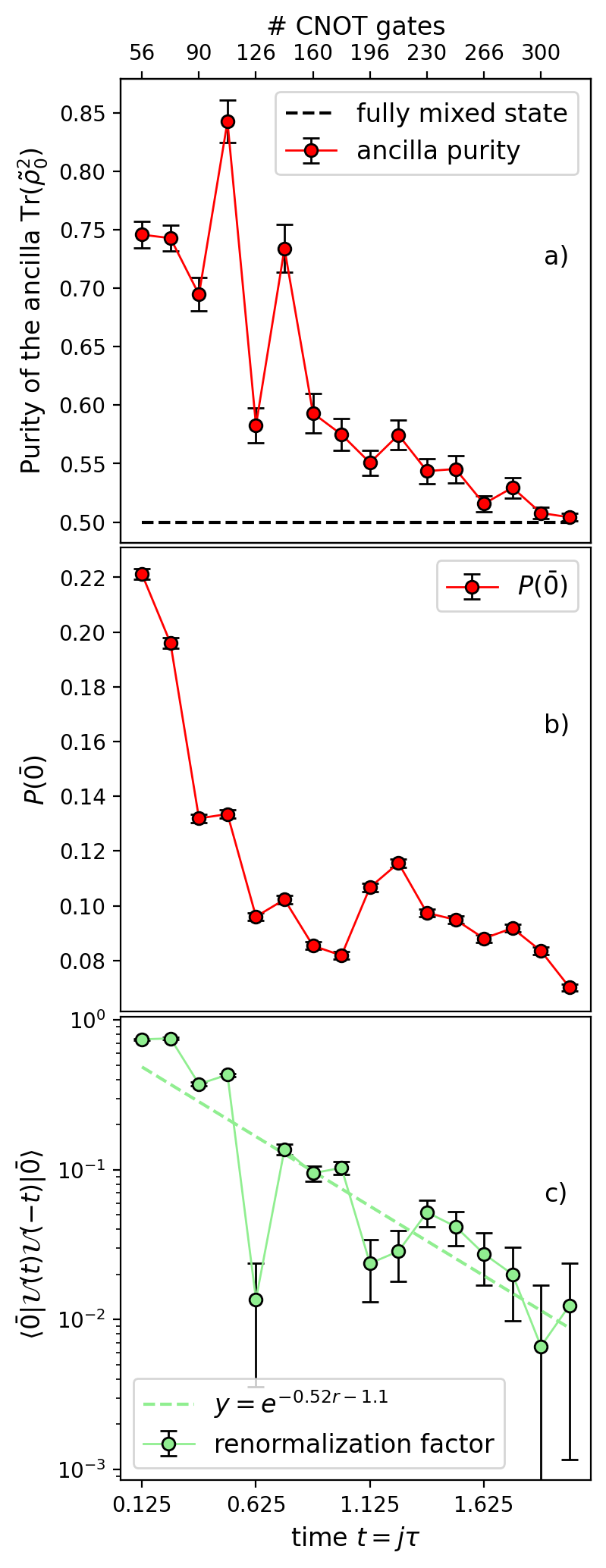}
    \caption{\textbf{Analysis} of the success of the mitigation strategies. Panel a) shows $\Tr{\tilde{\rho}_0^2}$ as a function of time and number of CNOT gates (top axis). The purity converges to the black dashed line at 0.5, which represents the purity of a fully mixed state. Panel b) displays the success probability. Panel c) exhibits the renormalization factor together with its exponential fit as a function of the number of Trotter steps $r$. The errorbars correspond to one sigma computed via Bayesian inference. The number of CNOT gates are indicated in the top axis, while the time is indicated at the bottom.}
    \label{fig:analysis}
\end{figure}

With PEV, the results are accurate up to seven steps, i.e. 266 CNOT gates, and start to deviate after that point. We can understand the large fluctuations after the breaking point by looking at the purity of the ancilla qubit, shown in the panel a) of Fig.~\ref{fig:analysis}. 

The purity converges to $0.5$, which corresponds to the one of a fully mixed state. At that point, the state is randomly projected onto the two eigenstates, resulting in high variance and bias. Moreover, the purity serves as a useful tool to check the success of the method, even in a regime where no reference solutions are available. Hence, one can check if the purity is above 0.5, to obtain some insurance on the success of the protocol. The panel b) shows the success probability $P(\bar{0})$ of ending in the correct state. This is important as it quantifies the number of samples required to estimate the expectation value to a given accuracy. We observe that the probability stays above 0.08 for the whole experiment. Therefore, using $16\times s$ shots is sufficient to achieve $1/\sqrt{s}$ accuracy. The few off points can be explained by the fluctuations of the device. 

The ODR strategy is less reliable, as it breaks down only after two double steps, and has a larger variance, especially without Pauli twirling. This is expected since, when the noise level is high, the division with $(1-p)$ becomes unstable. To better understand this, we show the renormalization factor as a function of time (and of the number of Trotter steps $r$) in Fig.~\ref{fig:analysis} c). We first observe that it decays exponentially with the depth, making the sampling more expensive as the number of Trotter step increases. Moreover, we can understand the less reliable moments results since they correspond to times where the renormalization factor fall by an order of magnitude. 

In both cases, the raw data is close to zero, meaning that the state is almost fully mixed. Thus, the error mitigation is primordial to extract any useful information from the experiments. 

\section{Conclusion}
In summary, our investigation underscores the role of Fourier moments in comprehending physical properties, particularly their utility in computing response functions \cite{faster_spectral_roggero} or spectra \cite{L&T}. Although quantum computers naturally facilitate the acquisition of moments, the challenges persist for NISQ devices due to controlled time evolution, introducing an overhead in CNOT gates and posing difficulties to overcome the decoherence of the devices. In this study, we address these challenges by employing control reversal gates~\cite{CRG,QETU} to alleviate control overhead and adapting error mitigation strategies, specifically echo verification \cite{OBrien_verification} and operator decoherence renormalization \cite{mitigating_depolarising,self-mitigation,farrell2023scalable}. We tailor these strategies to the computation of moments, which is arguably a difficult task due to the controlled time evolution. 

Our experiments, conducted on a real IBM quantum computer with superconducting transmon qubits, reveal noteworthy findings. In the first experiment, utilizing a single double Trotter step for computing the first ten moments, both purified echo verification and operator decoherence renormalization yield accurate results consistent with noise-free simulations. Further, employing the fast forwarding factor from control reversal gates maintains small Trotter errors when using noise renormalization, up to the tenth moment. 
In the second experiment, where multiple Trotter steps are executed until the breaking point, purified echo verification emerges as the most effective strategy, providing accuracy up to 266 CNOT gates. Conversely, operator decoherence renormalization becomes unstable after 90 CNOT gates, particularly without Pauli twirling, highlighting an inherent challenge related to the strategy involving division by a potentially small renormalization factor.

Particular care is given to the optimization of the runs using various error suppression techniques such as twirling, dynamical decoupling, pulse efficient transpilation, and control/readout error mitigation. Importantly, all these techniques, including echo verification and operator decoherence renormalization, prove to be inherently cost-effective, demanding only a constant increase in the number of samples. In conclusion, the combination of purified echo verification with error suppression techniques emerges as a powerful approach for extracting Hamiltonian moments from noisy quantum devices, exhibiting a linear increase in sample complexity as opposed to the exponential overhead associated with probabilistic error cancellation \cite{complexity_QEM}. We hope these technique to be useful when going to a larger scale regime, which will be the focus of future work.

\begin{acknowledgments}
We acknowledge useful conversations about the ODR scheme with R. Lewis in the early development stages of the present work.
O.K. is funded by the University of Geneva through a Doc.Mobility fellowship. M.G. is supported by CERN through the CERN Quantum Technology Initiative. A.R. is funded by the European Union under Horizon Europe Program - Grant Agreement 101080086 — NeQST. Views and opinions expressed are however those of the authors only and do not necessarily reflect those of CERN, IBM, the European Union or European Climate, Infrastructure and Environment Executive Agency (CINEA). Neither CERN, the European Union nor the granting authority can be held responsible for them. We also acknowledge interactions supported by the DOE HEP QuantISED grant KA2401032.
This work is part of the activities of the Quantum Computing for High-Energy Physics (QC4HEP) working group. Access to the IBM Quantum Services was obtained through the IBM Quantum Hub at CERN.

\end{acknowledgments}
\bibliography{bibliography}
% \onecolumngrid
\appendix

\section{Error suppression }
\label{error_suppresion}
In this section, we provide more details on the error suppression strategies used to enhance the results of the experiments. 
\subsection{Control noise mitigation}
Classical control, including measurements, is an additional source of decoherence and can deteriorate the estimation of expectation values. To mitigate this effect, we average $\langle Z\rangle$ and $-\langle -Z\rangle$, as proposed in \cite{OBrien_verification}. This scheme can be easily implemented in practice by applying a Pauli $X$ gate before the control and does not increase the required resources, as both circuit only requires half of the shot budget. This scheme effectively mitigates classical control errors. 

\subsection{Measurement error mitigation}
Measurement on the quantum hardware are suspect to errors, which can be mitigated by calibrating the device. We proceed by measuring the $n$-qubit state $|0\rangle^{\otimes n}$ and $|1\rangle^{\otimes n}$ as proposed in Ref.~\cite{Nation_PRXQ}, and use them to build the confusion matrices
\begin{equation}
P_k=
\begin{pmatrix}
P^{(k)}_{0,0} &P^{(k)}_{0,1} \\P^{(k)}_{1,0} &P^{(k)}_{1,1}
\end{pmatrix},
\end{equation}
where, $P^{(k)}_{i,j}$ is the probability of the $k$-th qubit to be in state $j\in\{0,1\}$ while measured in state $i\in\{0,1\}$.
The measurements $\vec{M}^k$ of the qubit $k$ can then be corrected as 
\begin{equation}
    \vec{M}^k_{\text{corrected}} = (P_k)^{-1}\vec{M}^k.
\end{equation}

\subsection{Randomized compiling}
Randomized compiling (RC) is a protocol aiming at turning coherent noise into stochastic noise by averaging over random equivalent circuits \cite{Randomized_compiling}. Stochasticity can drastically reduce the number of unpredictable errors in the computations due to the interaction with the environment. This can be achieved by twirling two\hyp qubit gates with individual qubit rotation. In practice, we compute $T$ random twirled circuits using $s/T$ shots, where $s$ is the total shot budget, and obtain an equivalent statistical distribution by computing the union over all results. This scheme has minimum overhead which happens at the transpilation time. More specifically, we consider Pauli twirling, where the random single\hyp qubit gates are taken from the set of Pauli strings. In a nutshell, Pauli twirling effectively turns the noisy channel into a Pauli channel 
\begin{equation}
\mathcal{E}(\rho) = \sum_{P\in \mathcal{P}^{\otimes n}} c_P P\rho P^\dagger,
\end{equation}
where $\mathcal{P} = \{\mathbb{1}, X, Y, Z\}$ is the set of Pauli matrices and $c_P$ the relative error due to $P$. We note that ways to compute smaller Pauli sets have been proposed in Ref.~\cite{Pauli_Twirling}.  
Tailoring coherent errors into stochastic Pauli noise has several significant advantages, as explored in Ref.~\cite{Randomized_compiling}: 
\begin{enumerate} \item The off\hyp diagonal terms in the error channel coming from coherent errors is suppressed, leading to smaller error rate. \\ 
\item Stochastic Pauli error only grows linearly with the circuit depth, in contrast to coherent noise accumulating up to quadratically with the circuit's depth (in the small error limit). Hence RC stabilizes the noise by preventing noise accumulation. Effectively, the noise accumulation under RC behaves similarly to a random walk and thus is quadratically slower. \\
\item RC is expected to work in symbiosis with ODR, since it turns the noise model into a depolarising channel, which can be mitigated using ODR. 
\end{enumerate}

\begin{figure*}
    \centering
    \includegraphics[scale=0.8]{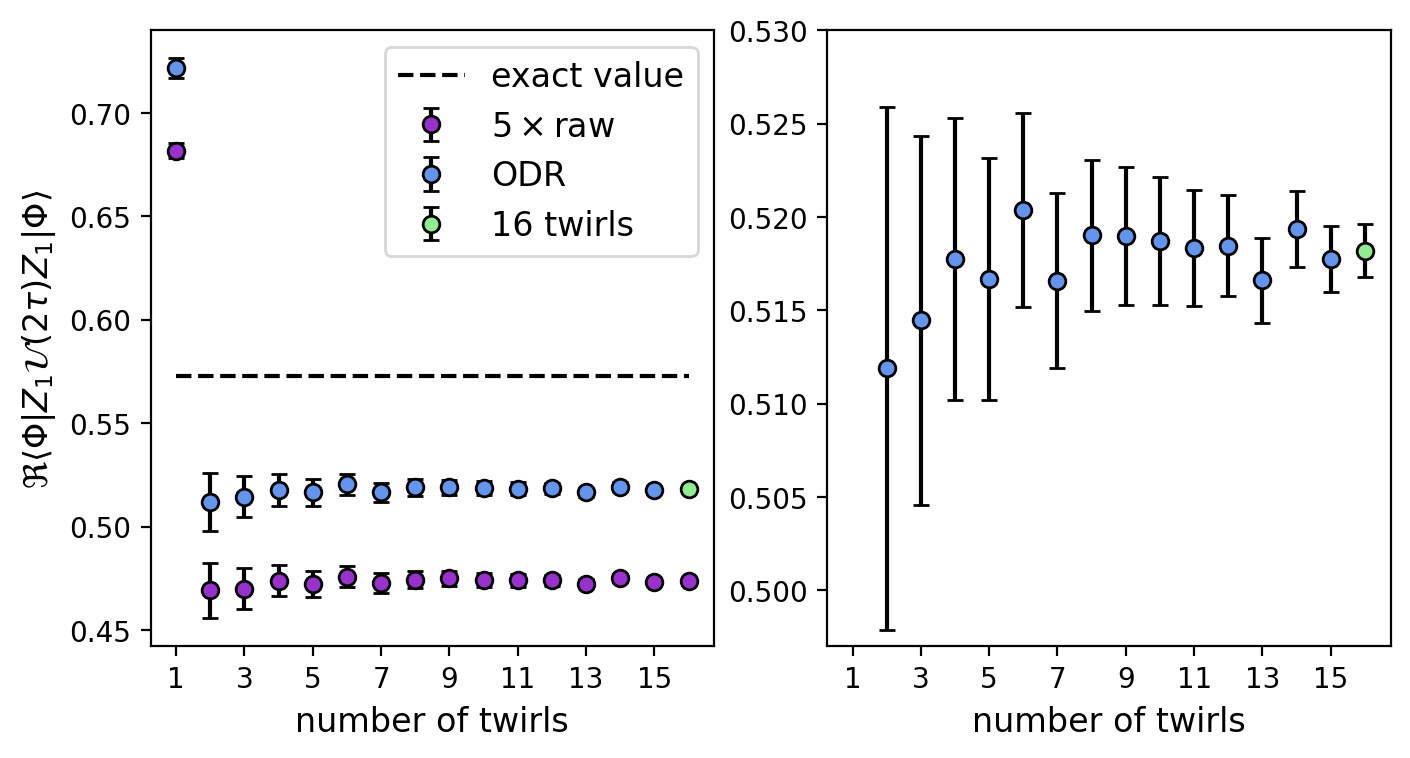}
    \caption{\textbf{Effect of twirling:} the expectation values are shown as a function of the number of twirls. The results with ODR are shown in blue, with the last point in green denotes the final result appearing in Fig.~\ref{fig:step_NR}. The raw data (scaled by a factor of 5 for better visibility) is shown in violet and the target value in a dashed black line. The second panel displays the same data but zoomed in to enhance the effect of the twirls. }
    \label{fig:twirls}
\end{figure*}

To better understand the effect of twirling, we consider the computation of the second moment with ODR as a function number of twirls, see Fig.~\ref{fig:twirls}. The fist panel shows the expectation values as function of the number of twirls. The results with ODR are shown in blue, with the last point in green denotes the final result appearing in Fig.~\ref{fig:step_NR}. The raw data (scaled by a factor of 5 for better visibility) is shown in violet and the target value in a dashed black line. The second panel displays the same data but zoomed in to enhance the effect of the twirls. We make two important observations: results with only one twirl, i.e. the original circuit, exhibit a larger bias and the effect of twirling seems to stabilize after ten twirls. 
 
\subsection{Dynamical decoupling}
Dynamical decoupling \cite{DD_Lloyd} is expected to increase the coherence time of quantum devices by taking advantage of short, time\hyp dependant, control modulation when the qubits are idling. Hence, idling qubits often decohere more easily, which can be mitigated by applying a sequence of gates that makes up the identity. For example, we can apply XX, YY, XY-2 or XY-8 gate sequences \cite{sequence_DD,sequence_DD2,DD_Lloyd}, as benchmarked in Ref. \cite{robust_DD}. It is important that the pulses sequences, e.g., the delay between pulses and their relative phases, are optimized for the specific devices used for the computation, e.g. using genetic algorithms \cite{genetic_DD}. In our experience, XY-8 pulse sequences seem to be the most effective.  

\subsection{Pulse efficient transpilation}
The native gate set of a specific quantum device can be different from the one used to write the quantum circuit. Hence, even if we use CNOT gates, the native two\hyp qubit gate of ibm\_kolkata is the cross resonance gate, equivalent to $\exp{-i\theta X\otimes Z}$. Therefore, we can rewrite some operations, notably the $\exp{-i\theta Z\otimes Z}$ gates using a cross resonance gate and some qubit rotations. These operations can often be implemented by pulses with shorter duration, thus slightly increasing the coherence time of the device \cite{pulse-efficient}. 

\section{Propagation of statistical uncertainties}
\label{bayesian}
Considering the multiple (non\hyp linear) operations involved in the processing of the raw data, it is difficult to estimate the related uncertainties. As described in Ref.~\cite[App. B]{PRD_neutrino}, a Bayesian strategy is used to infer the expectation values. The main idea is to use Bayes theorem to generate an arbitrary number of experiments, compatible with the bare data, and compute their variance. The procedure is best described considering a single qubit, whose probability of obtaining $m$ measurement of the $|1\rangle$ state out of a total of $M$ trials is given by a binominal distribution
\begin{equation}
    P_b(m;p) = \binom{M}{m}p^m(1-p)^{(M-m)}.
\end{equation}
The probability $p$ of obtaining $|1\rangle$ can be inferred with Bayes theorem  
\begin{equation}
    P(p|m_i) = \frac{P(m_i|p)P(p)}{\int dq P(m_i|q)P(q)},
\end{equation}
which can be obtain in closed form using a beta prior 
\begin{equation}
    P_\beta(p;\alpha,\beta) = \frac{\Gamma(\alpha+\beta)}{\Gamma(\alpha)\Gamma(\beta)}p^{(\alpha-1)}(1-p)^{(\beta-1)}.
\end{equation}
Here, $\alpha$, $\beta>0$ are the parameters of the beta distribution. They are initialised uniformly at $\alpha_0 = \beta_0 = 1$, and updated after each measurement as 
\begin{equation}
    \alpha_i = \alpha_0 + m_i \,\,\,\,\,\,\,\, \beta_i = \beta_0 + M - m_i.
\end{equation}
Expectations values can finally be perform after this inference step with the following procedure 
\begin{enumerate}
    \item sample a value $p_k'$ from the posterior $P(p_k'|m_i)$.\\
    \item sample $L$ new measurements from the likelihood $P_b(m_k';p_k')$. \\
    \item compute expectations values by averages over the generate measurements as $\langle O \rangle=\frac{1}{L}\sum_{k=1}^L \langle O_k \rangle $.
\end{enumerate}
The generalisation to multiple qubits is straightforward using a multinominal distribution, whose prior is given in closed form by the Dirichlet distribution. 

\onecolumngrid

\section{Explicit circuits} 
\label{explicit circuit}
In the section, we present the explicit circuits, discuss the optimization decomposition as well as the implementation of the control reversal gates.
We recall that the Hamiltonian of interest, see Eq.~(\ref{eq:nuclattice}) reads
\begin{equation}
H=\alpha\sum_kX_k + \beta\left(Z_1Z_4+Z_2Z_3+\sum_{i<j<k}Z_iZ_jZ_k\right)\;.
\end{equation}
 Since, we aim at performing Hadamard test using CRG, we aim to find a gate that anti\hyp commutes with the Hamiltonian, and control it on the ancilla. We remark that the phase of the $X$ terms can be controlled using either a $Y$ or a $Z$ while the diagonal part needs an $X$ or a $Y$. With linear connectivity we can do the following
\begin{equation}
\Qcircuit @C=1em @R=1em {
	\lstick{1}&\gate{Z}&\gate{\alpha/2}&\gate{Z}&\qw&\multigate{3}{\begin{matrix}&Z_1Z_4\\&Z_1Z_2Z_4\\&Z_1Z_3Z_4\\&Z_2Z_3Z_4\\ \end{matrix}} &\qw&\multigate{3}{\begin{matrix}&Z_2Z_3\\&Z_1Z_2Z_3\\ \end{matrix}} &\qw&\gate{Z}&\gate{\alpha/2}&\gate{Z}&\qw\\
	\lstick{4}&\gate{Y}\qwx&\gate{\alpha/2}&\gate{Y}\qwx&\gate{Y}&\ghost{\begin{matrix}&Z_1Z_4\\&Z_1Z_2Z_4\\&Z_1Z_3Z_4\\&Z_2Z_3Z_4\\ \end{matrix} }&\qw&\ghost{\begin{matrix}&Z_2Z_3\\&Z_1Z_2Z_3\\ \end{matrix}} &\gate{Y}&\gate{Y}\qwx&\gate{\alpha/2}&\gate{Y}\qwx&\qw\\
	\lstick{3}&\gate{Z}\qwx&\gate{\alpha/2}&\gate{Z}\qwx&\qw&\ghost{\begin{matrix}&Z_1Z_4\\&Z_1Z_2Z_4\\&Z_1Z_3Z_4\\&Z_2Z_3Z_4\\ \end{matrix}}&\qw&\ghost{\begin{matrix}&Z_2Z_3\\&Z_1Z_2Z_3\\ \end{matrix}} &\qw\qwx&\gate{Z}\qwx&\gate{\alpha/2}&\gate{Z}\qwx&\qw\\
	\lstick{2}&\gate{Z}\qwx&\gate{\alpha/2}&\gate{Z}\qwx&\qw&\ghost{\begin{matrix}&Z_1Z_4\\&Z_1Z_2Z_4\\&Z_1Z_3Z_4\\&Z_2Z_3Z_4\\ \end{matrix}}&\gate{X}&\ghost{\begin{matrix}&Z_2Z_3\\&Z_1Z_2Z_3\\ \end{matrix}} &\gate{X}\qwx&\gate{Z}\qwx&\gate{\alpha/2}&\gate{Z}\qwx&\qw\\
	\lstick{a}&\ctrl{-1}&\qw&\ctrl{-1}&\ctrl{-3}&\qw&\ctrl{-1}&\qw&\ctrl{-1}&\ctrl{-1}&\qw&\ctrl{-1}&\qw\\
 },
\end{equation}
where the box with $\alpha/2$ is the $X$ rotation of angle $\alpha/2$ for the one body and the other multi\hyp qubit gates implement the rotations proportional to $\beta$. Since $Z$ commutes with the big boxes we can simplify this to
\begin{equation}
\Qcircuit @C=1em @R=1em {
	\lstick{1}&\gate{Z}&\gate{\alpha/2}&\qw&\multigate{3}{\begin{matrix}&Z_1Z_4\\&Z_1Z_2Z_4\\&Z_1Z_3Z_4\\&Z_2Z_3Z_4\\\end{matrix}} &\qw&\multigate{3}{\begin{matrix}&Z_2Z_3\\&Z_1Z_2Z_3\\\end{matrix}} &\qw&\gate{\alpha/2}&\gate{Z}&\qw\\
	\lstick{4}&\gate{Y}\qwx&\gate{\alpha/2}&\qw&\ghost{\begin{matrix}&Z_1Z_4\\&Z_1Z_2Z_4\\&Z_1Z_3Z_4\\&Z_2Z_3Z_4\\\end{matrix} }&\qw&\ghost{\begin{matrix}&Z_2Z_3\\&Z_1Z_2Z_3\\\end{matrix}} &\qw&\gate{\alpha/2}&\gate{Y}\qwx&\qw\\
	\lstick{3}&\gate{Z}\qwx&\gate{\alpha/2}&\qw&\ghost{\begin{matrix}&Z_1Z_4\\&Z_1Z_2Z_4\\&Z_1Z_3Z_4\\&Z_2Z_3Z_4\\\end{matrix}}&\qw&\ghost{\begin{matrix}&Z_2Z_3\\&Z_1Z_2Z_3\\\end{matrix}} &\qw&\gate{\alpha/2}&\gate{Z}\qwx&\qw\\
	\lstick{2}&\gate{Z}\qwx&\gate{\alpha/2}&\qw&\ghost{\begin{matrix}&Z_1Z_4\\&Z_1Z_2Z_4\\&Z_1Z_3Z_4\\&Z_2Z_3Z_4\\\end{matrix}}&\gate{ZX}&\ghost{\begin{matrix}&Z_2Z_3\\&Z_1Z_2Z_3\\\end{matrix}} &\gate{XZ}&\gate{\alpha/2}&\gate{Z}\qwx&\qw\\
	\lstick{a}&\ctrl{-1}&\qw&\qw&\qw&\ctrl{-1}&\qw&\ctrl{-1}&\qw&\ctrl{-1}&\qw \\
 }\,\,\,,
\end{equation}
Now, since $XZ=-ZX$ we have
\begin{equation}
\Qcircuit @C=1em @R=1em {
	\lstick{1}&\gate{Z}&\gate{\alpha/2}&\qw&\multigate{3}{\begin{matrix}&Z_1Z_4\\&Z_1Z_2Z_4\\&Z_1Z_3Z_4\\&Z_2Z_3Z_4\\\end{matrix}} &\qw&\multigate{3}{\begin{matrix}&Z_2Z_3\\&Z_1Z_2Z_3\\\end{matrix}} &\qw&\gate{\alpha/2}&\gate{Z}&\qw\\
	\lstick{4}&\gate{Y}\qwx&\gate{\alpha/2}&\qw&\ghost{\begin{matrix}&Z_1Z_4\\&Z_1Z_2Z_4\\&Z_1Z_3Z_4\\&Z_2Z_3Z_4\\\end{matrix} }&\qw&\ghost{\begin{matrix}&Z_2Z_3\\&Z_1Z_2Z_3\\\end{matrix}} &\qw&\gate{\alpha/2}&\gate{Y}\qwx&\qw\\
	\lstick{3}&\gate{Z}\qwx&\gate{\alpha/2}&\qw&\ghost{\begin{matrix}&Z_1Z_4\\&Z_1Z_2Z_4\\&Z_1Z_3Z_4\\&Z_2Z_3Z_4\\\end{matrix}}&\qw&\ghost{\begin{matrix}&Z_2Z_3\\&Z_1Z_2Z_3\\\end{matrix}} &\qw&\gate{\alpha/2}&\gate{Z}\qwx&\qw\\
	\lstick{2}&\gate{Z}\qwx&\gate{\alpha/2}&\qw&\ghost{\begin{matrix}&Z_1Z_4\\&Z_1Z_2Z_4\\&Z_1Z_3Z_4\\&Z_2Z_3Z_4\\\end{matrix}}&\gate{XZ}&\ghost{\begin{matrix}&Z_2Z_3\\&Z_1Z_2Z_3\\\end{matrix}} &\gate{ZX}&\gate{\alpha/2}&\gate{Z}\qwx&\qw\\
	\lstick{a}&\ctrl{-1}&\qw&\gate{Z}&\qw&\ctrl{-1}&\gate{Z}&\ctrl{-1}&\qw&\ctrl{-1}&\qw\\
    }\,\,\,.
\end{equation}
Which simplifies to
\begin{equation}
\label{circ:R}
\Qcircuit @C=1em @R=1em {
	\lstick{1}&\gate{Z}&\gate{\alpha/2}&\multigate{3}{\begin{matrix}&Z_1Z_4\\&Z_1Z_2Z_4\\&Z_1Z_3Z_4\\&Z_2Z_3Z_4\\\end{matrix}} &\qw&\multigate{3}{\begin{matrix}&Z_2Z_3\\&Z_1Z_2Z_3\\\end{matrix}} &\qw&\gate{\alpha/2}&\gate{Z}&\qw\\
	\lstick{4}&\gate{Y}\qwx&\gate{\alpha/2}&\ghost{\begin{matrix}&Z_1Z_4\\&Z_1Z_2Z_4\\&Z_1Z_3Z_4\\&Z_2Z_3Z_4\\\end{matrix} }&\qw&\ghost{\begin{matrix}&Z_2Z_3\\&Z_1Z_2Z_3\\\end{matrix}} &\qw&\gate{\alpha/2}&\gate{Y}\qwx&\qw\\
	\lstick{3}&\gate{Z}\qwx&\gate{\alpha/2}&\ghost{\begin{matrix}&Z_1Z_4\\&Z_1Z_2Z_4\\&Z_1Z_3Z_4\\&Z_2Z_3Z_4\\\end{matrix}}&\qw&\ghost{\begin{matrix}&Z_2Z_3\\&Z_1Z_2Z_3\\\end{matrix}} &\qw&\gate{\alpha/2}&\gate{Z}\qwx&\qw\\
	\lstick{2}&\gate{Z}\qwx&\gate{\alpha/2}&\ghost{\begin{matrix}&Z_1Z_4\\&Z_1Z_2Z_4\\&Z_1Z_3Z_4\\&Z_2Z_3Z_4\\\end{matrix}}&\gate{X}&\ghost{\begin{matrix}&Z_2Z_3\\&Z_1Z_2Z_3\\\end{matrix}} &\gate{X}&\gate{\alpha/2}&\gate{Z}\qwx&\qw\\
	\lstick{a}&\ctrl{-1}&\qw&\qw&\ctrl{-1}&\qw&\ctrl{-1}&\qw&\ctrl{-1}&\qw\\
}\,\,\,.
\end{equation}
Note that the controlled Pauli at the edges need to be done only twice for any number of steps. These can be done in linear connectivity with seven CNOT gates
\begin{equation}
\Qcircuit @C=1em @R=1em {
	\lstick{1}&\gate{Z}&\qw\\
	\lstick{4}&\gate{Y}\qwx&\qw\\
	\lstick{3}&\gate{Z}\qwx&\qw\\
	\lstick{2}&\gate{Z}\qwx&\qw\\
	\lstick{a}&\ctrl{-1}&\qw\\
}\quad\quad=\quad\quad\Qcircuit @C=1em @R=1em {
	\lstick{1}&\gate{H}           &\targ    &\qw      &\qw      &\qw      &\qw      &\qw      &\targ    &\gate{H}\\
	\lstick{4}&\gate{S^\dagger H }&\ctrl{-1}&\targ    &\qw      &\qw      &\qw      &\targ    &\ctrl{-1}&\gate{HS }\\
	\lstick{3}&\gate{H}           &\qw      &\ctrl{-1}&\targ    &\qw      &\targ    &\ctrl{-1}&\qw      &\gate{H}\\
	\lstick{2}&\gate{H}           &\qw      &\qw      &\ctrl{-1}&\targ    &\ctrl{-1}&\qw      &\qw      &\gate{H}\\
	\lstick{a}&\qw                &\qw      &\qw      &\qw      &\ctrl{-1}&\qw      &\qw      &\qw      &\qw\\
}\,\,\,.
\end{equation}
Note also that the three Hadamard gates on the right can be moved past the $X$ rotations turning them into simpler $Z$ rotations. Now for the interaction parts, we start with three terms containing $Z_1Z_4$ and store their parity in qubit $2$ for the two three-body terms, we get
\begin{equation}
\Qcircuit @C=1em @R=1em {
	\lstick{1}&\multigate{3}{\begin{matrix}&Z_1Z_4\\&Z_1Z_2Z_4\\&Z_1Z_3Z_4\\\end{matrix}} &\qw\\
	\lstick{4}&\ghost{\begin{matrix}&Z_1Z_4\\&Z_1Z_2Z_4\\&Z_1Z_3Z_4\\\end{matrix} }&\qw\\
	\lstick{3}&\ghost{\begin{matrix}&Z_1Z_4\\&Z_1Z_2Z_4\\&Z_1Z_3Z_4\\\end{matrix}}&\qw\\
	\lstick{2}&\ghost{\begin{matrix}&Z_1Z_4\\&Z_1Z_2Z_4\\&Z_1Z_3Z_4\\\end{matrix}}&\qw\\
}\quad\quad=\quad\quad\Qcircuit @C=1em @R=1em {
	\lstick{1}&\ctrl{1}&\qw     &\qw         &\qw     &\ctrl{1}&\qw\\
	\lstick{4}&\targ   &\ctrl{2}&\gate{\beta}&\ctrl{2}&\targ   &\qw\\
	\lstick{3}&\qw     &\targ   &\gate{\beta}&\targ   &\qw     &\qw\\
	\lstick{2}&\qw     &\targ   &\gate{\beta}&\targ   &\qw     &\qw\\
}\,\,\,.
\end{equation}
so six CNOT gates with all-to-all connectivity (four can be removed using non-local $R_{ZX}$ gates). With linear connectivity this takes only two more CNOT gates
\begin{equation}
\label{circ:3block}
\Qcircuit @C=1em @R=1em {
	\lstick{1}&\multigate{3}{\begin{matrix}&Z_1Z_4\\&Z_1Z_2Z_4\\&Z_1Z_3Z_4\\\end{matrix}} &\qw\\
	\lstick{4}&\ghost{\begin{matrix}&Z_1Z_4\\&Z_1Z_2Z_4\\&Z_1Z_3Z_4\\\end{matrix} }&\qw\\
	\lstick{3}&\ghost{\begin{matrix}&Z_1Z_4\\&Z_1Z_2Z_4\\&Z_1Z_3Z_4\\\end{matrix}}&\qw\\
	\lstick{2}&\ghost{\begin{matrix}&Z_1Z_4\\&Z_1Z_2Z_4\\&Z_1Z_3Z_4\\\end{matrix}}&\qw\\
}\quad\quad=\quad\quad\Qcircuit @C=1em @R=1em {
	\lstick{1}&\ctrl{1}&\qw     &\qw     &\qw         &\qw     &\qw     &\ctrl{1}&\qw\\
	\lstick{4}&\targ   &\ctrl{1}&\qw     &\gate{\beta}&\qw     &\ctrl{1}&\targ   &\qw\\
	\lstick{3}&\ctrl{1}&\targ   &\ctrl{1}&\gate{\beta}&\ctrl{1}&\targ   &\ctrl{1}&\qw\\
	\lstick{2}&\targ   &\qw     &\targ   &\gate{\beta}&\targ   &\qw     &\targ   &\qw\\
}\,\,\,.
\end{equation}
The $Z_2Z_3Z_4$ term can be done directly in linear connectivity with four CNOT gates (or two plus one $R_{ZX}$) as
\begin{equation}
\label{circ:1block}
\Qcircuit @C=1em @R=1em {
	\lstick{4}&\multigate{2}{\begin{matrix}&Z_2Z_3Z_4\\\end{matrix}} &\qw\\
	\lstick{3}&\ghost{\begin{matrix}&Z_2Z_3Z_4\\\end{matrix}}&\qw\\
	\lstick{2}&\ghost{\begin{matrix}&Z_2Z_3Z_4\\\end{matrix}}&\qw\\
}\quad\quad=\quad\quad\Qcircuit @C=1em @R=1em {
	\lstick{4}&\ctrl{1}&\qw     &\qw         &\qw     &\ctrl{1}&\qw\\
	\lstick{3}&\targ   &\ctrl{1}&\qw         &\ctrl{1}&\targ   &\qw\\
	\lstick{2}&\qw     &\targ   &\gate{\beta}&\targ   &\qw     &\qw\\
}\,\,\,.
\end{equation}
Finally, we do the last one as in Eq.~\eqref{circ:3block} but reversed upside down and do only two rotations. Note that no explicit SWAP gates have been used.

\end{document}